\documentclass[fleqn,usenatbib]{mnras}

\usepackage{newtxtext,newtxmath}

\usepackage[T1]{fontenc}

\DeclareRobustCommand{\VAN}[3]{#2}
\let\VANthebibliography\thebibliography
\def\thebibliography{\DeclareRobustCommand{\VAN}[3]{##3}\VANthebibliography}

\usepackage{graphicx}	
\usepackage{amsmath}	
\usepackage{xspace}
\usepackage{xcolor}

\newcommand{\msun}{\ensuremath{M_{\sun}}}  
\newcommand{\kms}{\ensuremath{\rm km\,s^{-1}}}  

\newcommand{\sigmaoned}{\ensuremath{\sigma_{\rm 1d}}\xspace}
\newcommand{\rhalf}{\ensuremath{R_{\rm h}}\xspace}
\newcommand{\score}{ \ensuremath{{\cal{D}}}\xspace}
\newcommand{\tage}{\ensuremath{t_{\rm age}\xspace}}
\newcommand{\tdyn}{\ensuremath{t_{\rm dyn}\xspace}}
\newcommand{\tdyntrue}{\ensuremath{t_{\rm dyn,true}\xspace}}
\newcommand{\tstar}{\ensuremath{t_{\rm *}\xspace}}
\newcommand{\deltaage}{\ensuremath{\Delta\tage\xspace}}
\newcommand{\mtot}{\ensuremath{M_{\rm tot}}\xspace}
\newcommand{\mcl}{\ensuremath{M_{0}}\xspace}
\newcommand{\rcl}{\ensuremath{R_{\rm cloud}}\xspace}
\newcommand{\alphaturb}{\ensuremath{\alpha_{\rm turb}}\xspace}
\newcommand{\alphavir}{\ensuremath{\alpha_{\rm vir}}\xspace}

\newcommand{\dmed}{\ensuremath{\widetilde{d_{\rm *}}}\xspace}

\defcitealias{Kerr2021}{SPYGLASS-I}
\defcitealias{Kerr2022b}{SPYGLASS-II}

\newcommand{\starforge}{\textsc{STARFORGE}\xspace} 
\newcommand{\nbodycode}{\textsc{nbody7++GPU}\xspace} 
\newcommand{\fiducial}{\texttt{fiducial}\xspace}
\newcommand{\rsmall}{\texttt{R3}\xspace}

\newcommand{\alphaL}{\texttt{alpha1}\xspace}
\newcommand{\alphaH}{\texttt{alpha4}\xspace}

\newcommand{\Bten}{\texttt{Bx10}\xspace}
\newcommand{\Bhundred}{\texttt{Bx100}\xspace}

\title[Stellar populations in STARFORGE II]{Stellar populations in STARFORGE II: Comparison with observations}

\author[J.\ P. Farias et al.]{
        Juan P.\ Farias ,$^{1}$\thanks{E-mail: juan.farias@austin.utexas.edu}
        Stella S.\ R.\ Offner,$^{1}$
        Ronan Kerr,$^{2}$
        Michael Y. Grudi\'c$^{3}$
\\
$^{1}$Department of Astronomy, University of Texas at Austin, TX 78712, USA \\
$^{2}$Dunlap Institute for Astronomy \& Astrophysics, University of Toronto,
Toronto, ON M5S 3H4, Canada\\
$^{3}$Flatiron Institute, Center for Computational Astrophysics, 162 5th Ave, New
York, NY 10010, USA\\
}

\date{Accepted XXX. Received YYY; in original form ZZZ}

\pubyear{2025}

\begin{document}
\label{firstpage}
\pagerange{\pageref{firstpage}--\pageref{lastpage}}
\maketitle

\begin{abstract}
Recent studies suggest that most star-forming regions in our Galaxy form stellar associations rather than bound clusters. We analyse models from the STARFORGE simulation suite, a set of magneto-hydrodynamical simulations that include all key stellar feedback and radiative processes following star formation through cloud dispersal. We create synthetic observations by introducing observational biases such as random spurious measurements, unresolved binaries, and photometric sensitivity. These biases affect the measurement of the group mass, size, and velocity dispersion, introducing uncertainties of up to 100\%, with accuracy improving as the number of system members increases. Furthermore, models favouring the formation of groups around massive stars were the most affected by observational biases, as massive stars contribute a larger fraction of the group mass and are often missing from astrometric surveys like Gaia. We compare the simulations to the Cepheus Far North (CFN) region, and show that CFN groups may have formed in a low-density environment similar to those modelled in STARFORGE but with massive stars not located preferentially in groups. We also question the effectiveness of the kinematic traceback method, showing that it is accurate within 20\% only for certain associations with actual virial parameters above 2. However, observational biases can artificially raise the virial parameter by up to a factor ten, making it difficult to evaluate the reliability of the traceback age. Additionally, since stars continue to form during the dispersal of the parent cloud, we find no relation between the stellar-dynamical age difference and the length of the embedded phase. 

\end{abstract}




\section{Introduction}
Stars typically form in groups from dozens to thousands of stars
\citep{BT1987,Lada2003}. Some of these groups rapidly disperse, while others
remain gravitationally bound, moving and evolving together
\citep{Krumholz2020,Adamo2020a}. These bound groups are often referred to as
``star clusters'' often defined as groups of stars that formed together and are
able to remain relatively dense over long periods of time.

These aggregations of stars contrast visibly with the surrounding field stars,
allowing photometric studies to determine properties such as size and age
\citep[e.g.,][]{Kharchenko2013}.

Because star clusters are bound, their members experience many dynamical
interactions both during and after formation, which modify their initial
configurations \citep{BT1987}. In addition, information regarding their formation
is obscured by degeneracies, whereby their present-day properties and kinematics
can be reproduced by different sets of initial conditions
\citep[e.g.,][]{Pijloo2015,Wang2021}.

The formation of bound star clusters is currently a matter of debate
\citep[see][for a review of the proposed models]{Krumholz2020}. A critical stage in
this process is how stars emerge from their natal gas while retaining enough
members to remain bound and dense over time
\citep{Hills1980,Kroupa2001,Baumgardt2007}. A cluster retains more stars if its
stellar mass constitutes a large fraction of the total system mass—i.e., it was
formed in molecular cloud with a high star formation efficiency (SFE) or if gas
leaves the system slowly, allowing the stellar orbits to adjust to the diminishing
gravitational potential \citep{Baumgardt2007,Adams2000,Smith2013}. However, growing
evidence suggests that star-forming regions generally do not meet these conditions,
implying that most star formation events produce associations, i.e., groups that
are not gravitationally bound,  rather than star clusters
\citep{Lada2003,Kruijssen2012b,Ward2020,Dinnbier2022,Wright2022}.

Stellar feedback is a key factor that regulates the efficiency of star formation
and determines how quickly gas is dispersed from the birth environment, thereby
influencing whether the stars remain bound \citep{Krause2020}. Ionizing radiation
and stellar winds from massive stars heat and inject momentum into the surrounding
gas, generating an expanding bubble around the newly formed stars
\citep{Dale2005,Haid2018,Grudic2022}. While protostellar jets regulate stellar masses
\citep{Guszejnov2021}, they also inject momentum, potentially driving turbulence on
small scales \cite[e.g.,][]{Nakamura2007,OffnerChaban2017}, maintaining low star
formation rates (SFR) \citep{Murray2018,Guszejnov2022}, and with it, preventing
the stellar dominance in the gravitational potential of the region. Ultimately, one
of the massive stars might explode as a supernova, clearing out any remaining gas
not yet removed by the previous processes \citep{Dinnbier2020}. While supernova
explosions might trigger some star formation in the expanding bubble \citep[see
e.g.][]{Dale2013a}, by this point, most star formation within the molecular cloud
is finished \citep{Grudic2022,Guszejnov2022}.  Overall, the gas dissipation process
by feedback lasts a few dynamical times \citep{Dinnbier2020,Farias2023a}.

It is now possible to model this process self-consistently, from the collapse of
the parent cloud to gas dispersal, via magneto-hydrodynamical (MHD) simulations.
The \starforge\ simulations \citep{Grudic2021,Grudic2022}, which we adopt in this
study, include all the key stellar feedback processes while still resolving
individual star formation down to $M_* \approx 0.1 \msun$. These simulations
produce an overall SFE of less than 20\% and do not form large stable bound
clusters but rather associations along with several small expanding groups that may
or may not contain bound stars \citep{Guszejnov2022,Farias2023a}. Most of these
groups continue expanding and dissolve within 20 Myr, while only a small fraction
of stars (below 40\%) remain part of a bound sub-cluster \citep{Farias2023a}.
However, most of these sub-clusters reach surface densities below 1~star/pc$^2$ in
less than 25 Myr. Provided that the \starforge\ simulations accurately represent
typical Milky Way clouds, these results suggest that the majority of star formation
events end in unbound expanding associations. 

With the advent of the Gaia mission, astrometric measurements have achieved
accuracies down to 0.1~mas/yr \citep{GaiaDR2}, which represents a ten-fold
improvement over the previous HIPPARCOS mission \citep{Urban1998}. This accuracy
has enabled a range of studies that identify stellar groups according to their
coherent motions, thereby increasing the discovery rate of low-density associations
\citep{Kuhn2019,Kerr2021,Cantat-Gaudin2019,Chemel2022}. Unbound stellar
associations are particularly important populations, because, in contrast to bound
stellar clusters, their kinematics are expected not to have changed significantly
since their formation. Consequently, the present-day stellar motions potentially
reflect the kinematic information of their parent cloud, thereby providing
important constraints on the star formation process, such as the velocity
dispersion at birth \citep{Wright2020}.

Kinematic accuracy has also enabled ``trace back" studies, which reverse the
trajectories of the individual stars to predict their evolution, ultimately
providing estimates of the time when the association began its expansion, i.e., the
\emph{dynamical age} \citep{Kerr2022b,Miret-Roig2024}. However, there are a number
of uncertainties that affect these analyses, including systematic biases in the
stellar ages, association membership, and completeness. Furthermore, even if these
uncertainties are small, a linear interpolation of the stellar velocities neglects
the the potentially complex history of associations and their star formation
history.

In this paper, we extend the work of \citet{Farias2023a}, which analysed the
\starforge\ stellar distributions, to investigate the implications of observational
uncertainties and simplifying assumptions on the inferred group properties and
traceback analyses. We perform this analysis by creating synthetic observations
that include stellar incompleteness and survey biases and use observational
techniques to infer the age, formation timescale and dynamical states. We test the
accuracy of such techniques, estimate the systematic biases, and discuss the
implications for Gaia studies of stellar groups. 

\section{Methods}

In this paper, we continue the work presented in \cite{Farias2023a}, where we
post-evolve stellar systems formed from clouds with solar neighborhood conditions
up to 200 Myr after their expansion. In this section, we summarize the key features
and relevant parameters of the simulations that give birth to these stellar
systems, and the code and methods we use to model their post-formation phase,
including the identification and analysis of subclusters within these stellar
regions. For further details of the models, we refer to \citep{Farias2023a} and
references therein. 

\subsection{\starforge Methods}

Stellar systems in this work are born within the \starforge\ simulations suite
\citep{Grudic2021} developed to model Giant Molecular Clouds (GMCs) from their
initial collapse to the dispersal of gas, including all key stellar feedback
processes, i.e.\ protostellar outflows, stellar winds, radiation pressure,
photoionization, and supernovae. The simulations are performed with the GIZMO code,
which uses a Lagrangian meshless finite-mass (MFM) method to solve the
magnetohydrodynamics (MHD) equations under the ideal MHD approximation
\citep{Hopkins2015,Hopkins2016a}. The self-gravity of the gas is modelled with an
improved version of the Barnes \& Hut tree algorithm \citep{Springel2005a}. The
code also includes a thermo-chemistry module from \cite{fire3} that models cooling
and heating mechanisms, including recombination, thermal bremsstrahlung, metal
lines, molecular lines, fine structure, and dust collisional processes within the
temperature range of $T = 2.7-10^{10}$\,K \citep[see][for details]{fire3}.

Protostars are represented by sink particles with a sink radius of 18\,AU, that is
also used as the gravitational softening length by the fourth-order Hermite N-body
integrator that follows their trajectories. The protostars follow a sub-grid model
for stellar evolution based on \cite{Offner2009b}. 

See \cite{Grudic2021} for a detailed description of the code features and
implementation.

\subsection{\starforge models}

We adopt the same simulation suite as \cite{Farias2023a} to serve as a basis for
our study. We select a subset of simulations from \cite{Grudic2022} and
\cite{Guszejnov2022a} to follow the evolution of the stellar complexes after gas
removal, starting from a fiducial model simulation, which we briefly describe here.
We refer the reader to \citet{Grudic2021} for the detailed description of the
simulations.

The standard molecular cloud modelled by \starforge\ (\fiducial\ model) consists of
an initially uniform sphere of gas with $\mcl= 20,000~\msun$ contained within a
radius of $\rcl=10$\,pc and with an initial temperature $T=10$~K. The cloud is
embedded in a warm $T=10^4$~K, diffuse medium 1000 times less dense than the
modelled cloud, such that the cloud and the ambient medium are roughly in thermal
pressure equilibrium. The initial dynamical state of the cloud is parameterised by
the \emph{turbulent virial parameter} (\alphaturb), which is defined as the ratio
between the cloud's kinetic and potential energy \citep{Bertoldi1992}. The internal
velocities are scaled to match a typical cloud in the Milky Way, i.e.,
$\alphaturb=2$ \citep{Larson1981,Chevance2022}. The initial turbulence of the cloud
is set by a Gaussian random velocity field with a power spectrum $E_k \propto
k^{-2}$ scaled to match \alphavir.

The simulated molecular clouds are magnetized where the strength of the fields is
scaled relative to the cloud's gravitational energy and parameterised as: 
\begin{eqnarray} 
        \mu &= & c_1 \sqrt{\frac{-E_{\rm grav}}{E_{\rm mag}}}, 
\end{eqnarray} 
where $E_{\rm grav}$ and $E_{\rm mag}$ are the gravitational and magnetic energies,
respectively, and $c_1\approx0.42$ is a normalisation constant such that $\mu= 1$
represents a critically stable homogeneous sphere in a uniform magnetic field
\citep{Mouschovias1976}. The fiducial simulation value is $\mu=4.2$, i.e., $E_{\rm
mag} = 0.1E_{\rm grav}$.

External radiation in the form of the interstellar radiation field (ISRF) is
included in the simulations assuming solar neighbourhood conditions
\citep{Draine2010}. Dust abundances in the clouds are assumed to have solar
metallicity with a dust-to-gas ratio of 0.01. 

In this paper, we also include simulations with variations of these initial
conditions, including low and high turbulent velocity field runs \alphaL
($\alphaturb=1$) and \alphaH\ ($\alphaturb=4$), respectively; increased magnetic
field strengths by 10 (\Bten) and 100 (\Bhundred), and a model \rsmall that has an
initial density ten times larger, i.e., setting $\rcl= 3$,pc.

\cite{Guszejnov2022a} demonstrated that the evolution of the fiducial cloud from
its initial global collapse and fragmentation results in the formation of stars
with an average SFE on the order of $\sim10\%$. Most stars form within 2 initial
free-fall times. After cloud collapse, when the distribution of newly formed stars
reaches the most compact configuration \citep[as shown in][]{Farias2023a}, stellar
feedback disperses the gas on a timescale of about half the initial freefall time
\citep{Farias2023a}. This rapid gas expulsion is too fast for the new stars to
adapt their orbits, and after this first collapse the system begins to expand. The
\starforge\ simulations finish after the first supernova explosion that, in the
\fiducial\ case, happens at about 9 Myr. At this point, the gas cloud is mostly
dispersed, and star formation is finished. However, we note that in general, these
simulations end a few Myr after the beginning of the stellar expansion, and
therefore they capture most of the initial expansion process that is important in
the analysis performed here.

\subsection{N-body simulations}
In \cite{Farias2023a} we followed the resulting \starforge\ stellar distributions for
an additional 200 Myr using the code \nbodycode \citep{Aarseth2003,Wang2015} with
no gas particles. 

\nbodycode is a fourth order Hermite integrator designed to efficiently model
the evolution of dense star clusters including high fractions of binaries and
higher order multiples. Together with an efficient time-stepping algorithm
(\emph{hierarchical block timesteps}) and sophisticated subroutines that solve
the trajectories of close encounters, binaries (termed \emph{KS
regularizations}) and higher order multiples \citep[\emph{chain
regularizations}, see ][for details]{Aarseth2003} it does not require the use of
a gravitational softening radius as other integrators use.

Given the change of numerical integration, from one scheme that uses gravitational
softening to another that does not, close binaries require a correction to their
orbital velocities. The softened gravity at close distances ($\lesssim 100$\,AU)
causes orbital velocities that are slower than in realty, resulting in orbits that
are artificially eccentric when gravity is no longer softened. We fix this by
correcting the orbits by choosing a new eccentricity drawn from a uniform
distribution \citep[see][for details]{Farias2023a}.

\subsection{Identification of groups}
As shown in our previous work, a natural outcome of turbulent fragmentation is that
each simulation does not form a homogeneous spherical star system, but rather forms
several sub-groups that may merge forming larger groups \citep{Guszejnov2022}.
While any initial substructure may be erased if the embedded phase lasts long
enough \citep{Smith2013,Farias2015}, the removal of gas occurs in less than a
free-fall time. Therefore, some substructure remains, and those groups separate
from each other as the region expands. In \cite{Farias2023a}, we identified several
groups formed in each simulation and analyze them independently. We developed a
method to identify systems based on an energy criterion, i.e., finding bound groups
of stars, which we refer to as \emph{star clusters}. We also identified stellar
systems based only on position using the Hierarchical Density-Based Spatial
Clustering of Applications with Noise (HDBSCAN) algorithm
\citep{Campello2013,McInnes2017,Malzer2020} implemented in
\textsc{Python}\footnote{https://pypi.org/project/hdbscan/}. 

Given a required minimum number of stars per group ($N_{\rm min}$) and a choice of
number of neighbours, $k$, required to define a neighbourhood radii, HDBSCAN finds
stars that are close to each other based on their Euclidean distance. It creates a
hierarchical tree of particles connected to each star within the neighbour radius.
Each branch of the tree represents a group that can be separated from the rest by
choosing a scale length. However, HDBSCAN does this algorithmically, by walking the
tree and selecting those groups that persist over several scales. We provide a full
description of this algorithm in \cite{Farias2023a} \citep[see also][for further
details]{Campello2013}.

In \cite{Farias2023a}, we further processed the selection to remove transient
groups, since small variations in the stellar distribution sometimes change the
membership selection and produces spurious, short-lived groups. We applied the
HDBSCAN algorithm to stars in 200 equally spaced times in the post-\starforge
evolution and recorded the group membership of each star. We then applied an
algorithm to match each star with its most robust and stable parent group,
reassigning its membership in snapshots where the star is assigned to another group
\citep[we refer to][ \S2.3.2 for a detailed description of this
algorithm]{Farias2023a}. In this way, we obtained more stable evolution of derived
quantities. Note that we performed the identification of groups with HDBSCAN directly 
on the simulation data. After the groups are well defined, we apply observational biases
in order to compare them with observations (see \S\ref{sec:matching}).

\subsection{Gas-free evolution}
Using the \nbodycode\ code, we modelled the gas-free phase of the new stellar
complexes for 200 Myr after the end of the \starforge\ simulations, i.e. the
gas-free phase. During this phase, the stellar distributions globally expand.
However, during the early expansion, some stellar groups condense and form bound
but expanding systems. These groups expand at a lower rate than the rest of the
region, but they reach surface densities below 1 star/pc$^2$ in about 25 Myr and,
in general, reach a stable surface density of around 0.1 stars/pc$^2$ that can be
retained for about a hundred Myr. 

\citet{Farias2023a} showed that these bound systems follow a characteristic
mass-size and mass-$\sigma^{\rm 3D}$ relation  that appears invariant to the cloud
initial parameters. These scaling relations, however, only hold for true bound
systems. For groups identified by the clustering algorithm rather than through an
energy criterion, the scaling relations vanish.

\subsection{Model cluster ages and traceback age}
\label{sec:agemethods}

There are a number of key timescales that characterize the stars and stellar
groups. Each star in the simulations has a known age (\tstar), which we define as
the time since the star first appeared in the simulation as a sink particle.
Therefore, it is possible to assign an age and age spread for each group to compare
with observations.

As shown in previous works \citep{Guszejnov2022,Farias2023a}, a single cloud
produces  several groups that may or may not merge and expand in different
directions. Each of these groups may contain stars formed at different stages of
the simulation. Consequently, the members of each group do not necessarily come
from the same region within the cloud. Therefore, we define the age of a group
(\tage) as the median age of its members. Each group is characterized by an age
spread (\deltaage), which we define as the standard deviation of group member ages.

In line with recent observational methods, we also compute the dynamical age of the
groups (\tdyn) \citep{Ducourant2014,Miret-Roig2018}. Observationally,
this time represents an age estimate of groups based on their current proper
motions. Stellar trajectories are traced back, identifying a time when stars were
at their smallest volume configuration. While \tdyn has been used as a
photometrically-independent estimation of the group age, it effectively estimates
the time since stellar groups began their expansion. 

We use a simple linear extrapolation of the stars' trajectory, which does not
account for the interactions between stars, since such processes are highly
uncertain in observed stellar associations. Following previous works
\citep[e.g.][]{Kerr2022b}, we consider the median mutual distance (\dmed) between
stars as a distance metric, since it does not require choosing a centre. The
dynamical age of a group is defined as the time when \dmed\ reaches a minimum
according to the linear traceback of the stellar orbits.

We use two methods to estimate dynamical ages. We expect to obtain the most
accurate results when utilizing all available group members. Therefore, we first
include all members but remove any binaries below 10,000 AU, as described in
\S\ref{sec:matching}. We refer to this dynamical age as $\tdyn^*$. We also estimate
dynamical ages by assuming the group is located a given distance from the Sun and
introducing the observational biases described in \S\ref{sec:matching} for the
chosen distance, while also removing unresolved binaries. We refer to this
dynamical age as $\tdyn$. In this way, we obtain two metrics: one that represents a
best case estimate, and the second, a method that reflects the primary
observational limitations.

Using the simulation history for each group, we also estimate the actual time when
the group begins expanding, \tdyntrue. While more accurate, this is not necessarily
simple, because member stars often form at different times, sometimes quite late in
the simulation, so it is not always clear where the expansion begins. Therefore, we
define the true dynamical age \tdyntrue as the time since the group members reach a
minimum \dmed and at least 50\% of their members are present.

\subsection{Matching model clusters and observed clusters}
\label{sec:matching}

\subsubsection{SPYGLASS and Cepheus Far North}
In this work we include observational biases in the measurements of the physical
properties of associations formed within \starforge. Most recent associations are
identified in data from the Gaia mission \citep{GaiaDR2}, which provides
unprecedented astrometric accuracy, including a robust multi-epoch photometric
system in conjunction with stellar evolution models such as 
PARSEC \citep{Bressan2012}, \cite{Baraffe2015}. Such models play an essential role
in identifying young populations of stars that may have formed together and
separating them from the older field star population. 

Despite the advances of Gaia, assigning ages to individual stars is complicated by
measurement uncertainties, reddening and other factors such as metallicity and
multiplicity \citep{Sullivan2021,Plotnikova2022}.
Consequently, most studies define associations via clustering algorithms that are
applied to the astrometry and kinematics in order to identify members that are
close to each other in phase-space. 

Along these lines, \cite{Kerr2021} developed a robust framework within the Stars
with Photometrically Young Gaia Luminosities Around the Solar System (SPYGLASS)
project. By combining Bayesian statistics, Gaia photometry, and PARSEC stellar
models, they surveyed the solar neighbourhood within 333 pc, distinguishing young
stars from the older field population and identifying over 30,000 probable young
stellar objects. They employed the HDBSCAN algorithm to classify candidates
into 27 associations, approximately half of which were largely unknown.

\citetalias{Kerr2022b} conducted an in-depth dynamical analysis of one of the
largest stellar associations in our neighbourhood, the Cepheus Far North
Association (CFN), which is located an average of 179 pc from the Sun. This work
expanded on the sample of 219 candidate members presented in \citetalias{Kerr2021},
identifying 549 candidate members spanning about 100~pc. They supplemented radial
velocity measurements from Gaia with literature sources (using SIMBAD and VIZIER).
Using HDBSCAN clustering in 5D space-transverse velocity coordinates,
\citetalias{Kerr2022b} divided CFN into 7 distinct subgroups. For each of these
groups, they computed total mass, velocity dispersion, half-mass radius, and
age. Additionally, they performed a 3-D dynamical traceback analysis, providing
the time that minimises the mutual relative distance between stars in a group,
which, in principle, indicates the moment each group began its expansion, \tdyn.

The seven identified CFN groups are relatively small, with masses ranging from 14
to 76 \msun~and half-mass radii (\rhalf) between 2 and 18 pc. The authors concluded
that these groups formed over a span of 10 Myr, with two spatially separated nodes
emerging between 16 and 26 Myr ago. This suggests a complex, prolonged star
formation event characterized by significant substructure.

\subsubsection{Matching procedure}
\label{sec:matchingprocedure}
\begin{table*}
\caption{
Properties of groups identified  in the STARFORGE simulations measured at 25 Myr. Properties with the ``obs" subscripts represent the values measured after placing the groups at a distance of 400\,pc, applying observational biases, and resampling 100 times. We report the median as the measured value with the 25 and 75 percentiles as errors. The tenth column, Age, reports the median age of the stars in the group. Columns 11, 12 and 13 show the traceback ages measured using all stars in the group ($t_\text{dyn}$), applying observational biases ($t^*_\text{dyn}$) and the true traceback age using simulation data ($t_\text{dyn,true}$).
}

\label{tab:data}
\begin{center}
		\begin{tabular}{|r|c|c|c|c|c|c|c|c|c|c|c|c|}
			
			Group  &  $M$  &  $M_\text{obs}$  &  $R_\text{h}$  &  $R_\text{h,obs}$  & $\sigma_\text{1d}$  &  $\sigma_\text{1d,obs}$  &  $\alpha_\text{vir}$  & $\alpha_\text{vir,obs}$  &  Age  &  $t_\text{dyn}$  &  $t^*_\text{dyn}$  &  $t_\text{dyn,true}$ \\ 
            
			&  [$\msun$]  &  [$\msun$]  &  [pc]  &  [pc]  & $[\kms]$  &  $[\kms]$  &    & & [Myr] &  [Myr]  & [Myr]   &  [Myr] \\ \hline
			
			\multicolumn{3}{l}{\texttt{fiducial}} & & & & & & & & & \\ 
            0 & 58.4   & $53^{+5}_{-6}$      & 25.4 & $25^{+1}_{-4}$ & 0.7 & $0.52\pm0.04$  & 24.48 & $15^{+4}_{-3}$      & 21.2 & 18.9 & $19.4^{+0.1}_{-0.5}$ & 19 \\
			
			1 & 1429.2 & $860\pm10$          & 38.9 & $38.6\pm0.3$   & 0.9 & $0.97^\pm0.01$ & 3.03 & $5.3^{+0.2}_{-0.1}$ & 21   & 19.4 & 19.4                 & 19.3 \\
			\hline\multicolumn{3}{l}{\texttt{R3}} & & & & & & & & & \\ 
            
			0 & 387.7  & $156\pm6$           & 1.9  & $2.20^{+0.16}_{-0.04}$ & 0.3 & $0.238^{+0.006}_{-0.005}$ & 0.040 & $0.1^{+0.009}_{-0.007}$ & 24.4 & 0 & 0 & 24.4 \\
			           
			1 & 208.8  & $133^{+5}_{-6}$     & 6    & $9.8^{+0.3}_{-1.0}$    & 0.2 & $0.186^{+0.007}_{-0.004}$ & 0.12  & $0.31^{+0.04}_{-0.03}$ & 24.3 & 11.7 & $10.7^{+0.1}_{-0.5}$ & 24.4 \\
			           
			2 & 128.9  & $32\pm2$            & 5    & $15.1^{+0.4}_{-0.2}$   & 0.4 & $0.37^{+0.01}_{-0.03}$    & 0.63  & $7\pm1$ & 24.5 & 25 & $25.0^{+0.1}_{-1.0}$ & 24.4 \\
			           
			3 & 339.6  & $67\pm3$            & 0.2  & $1.64\pm0.06$          & 0.3 & $0.312^{+0.009}_{-0.008}$ & 0.0070& $0.30^{+0.03}_{-0.02}$ & 24.4 & 0 & 0 & 24.4 \\
			           
			4 & 225.9  & $62^{+4}_{-7}$      & 13.6 & $9.6^{+4.0}_{-0.2}$    & 0.3 & $0.232^{+0.013}_{-0.008}$ & 0.52  & $1.1^{+0.4}_{-0.1}$ & 24.3 & 22.4 & $20^{+1}_{-2}$ & 24.4 \\
			           
            5 & 114.5  &$30.8^{+0.8}_{-6.7}$ & 1.9  & $1.95^{+0.07}_{-0.08}$ & 0.2 & $0.23\pm0.02$             & 0.12  & $0.46^{+0.28}_{-0.07}$ & 24.6 & 1.5 & $1.0\pm0.5$ & 24.5 \\
			           
			6 & 156.3  & $35^{+2}_{-3}$      & 0.5  & $1.9^{+0.4}_{-0.1}$    & 0.1 & $0.151^{+0.009}_{-0.008}$ & 0.0088& $0.16^{+0.03}_{-0.02}$ & 24.2 & 0 & 0 & 24.4 \\
		
			\hline\multicolumn{3}{l}{\texttt{alpha1}} & & & & & & & & & \\ 
			0 & 95.2   & $46^{+2}_{-3}$  & 6.6   & $9^{+1}_{-2}$         & 0.3 & $0.3^{+0.01}_{-0.02}$       & 0.65 & $1.8^{+0.5}_{-0.3}$ & 19.3 & 18.9 & 18.88                   & 18.5 \\
			
			1 & 1341.4 & $740\pm10$      & 9.5   & $10.9^{+0.4}_{-0.6}$  & 0.4 & $0.402^{+0.004}_{-0.003}$   & 0.15 & $0.29\pm0.01$       & 20.8 & 19.4 & $19.39^{+0.01}_{-0.51}$ & 20.1 \\
			
			2 & 70.1   & $25^{+1}_{-3}$  & 11.6  & $15^{+1}_{-3}$        & 0.4 & $0.42^{+0.06}_{-0.02}$      & 2.82 & $11^{+4}_{-2}$      & 21.6 & 17.3 & $14.8^{+0.5}_{-1.0}$    & 20.3 \\
			
			\hline\multicolumn{3}{l}{\texttt{alpha4}} & & & & & & & & & \\ 
			0 & 93.4   & $48\pm3$  & 8.5     & $14.5^{+0.3}_{-0.4}$  & 0.4 & $0.43\pm0.01$ & 1.54 & $6.7^{+0.8}_{-0.5}$   & 18.8 & 15.8 & $14.29^{+0.51}_{-0.01}$ & 18.1 \\
			
			1 & 407.7  & $298\pm7$ & 20.5    & $19.1\pm0.4$          & 0.5 & $0.54\pm0.01$ & 1.75 & $2.4\pm0.1$           & 19.2 & 15.8 & $15.82^{+0.01}_{-0.51}$ & 17.9 \\
			
			\hline\multicolumn{3}{l}{\texttt{Bx10}} & & & & & & & & & \\ 
			0 & 102.5  & $80\pm4$        & 13.3  & $15.9^{+0.2}_{-0.5}$  & 0.5 & $0.38\pm0.02$               & 3.35  & $3.6\pm0.4$           & 18.7  & 14.8  & $14.29^{+0.01}_{-0.51}$   & 17.6  \\
			
			1 & 110.8  & $38^{+1}_{-3}$  & 0.8   & $2.4\pm0.1$           & 0.2 & $0.22^{+0.01}_{-0.02}$      & 0.043 & $0.37\pm0.06$         & 19.5  & 3.6   & $3.1\pm0.5$               & 17    \\
			
			2 & 218.3  & $15^{+4}_{-5}$  & 11.2  & $10.3\pm0.2$          & 0.4 & $0.338^{+0.006}_{-0.005}$   & 0.80  & $0.96^{+0.05}_{-0.04}$& 17    & 15.8  & 15.31                     & 16    \\
			
			3 & 110.5  & $63\pm4$        & 6.5   & $8.0\pm0.4$           & 0.3 & $0.244^{+0.005}_{-0.008}$   & 0.46  & $0.94\pm0.09$         & 20.3  & 14.8  & 13.78                     & 18    \\
			
			4 & 77.9   & $44^{+1}_{-4}$  & 0.9   & $2.5^{+0.2}_{-0.1}$   & 0.2 & $0.186^{+0.003}_{-0.004}$   & 0.053 & $0.24\pm0.02$         & 20    & 7.7   & $11.2^{+1.5}_{-0.5}$      & 17.4  \\
			
            \hline\multicolumn{3}{l}{\texttt{Bx100}} & & & & & & & & & \\ 

			0 & 63.5   & $41^{+1}_{-2}$      & 3.9   & $6.5^{+0.2}_{-0.6}$       & 0.4 & $0.382^{+0.006}_{-0.008}$   & 1.11  & $2.9^{+0.3}_{-0.2}$   & 15.6  & 6.1   & $6.12 \pm 0.01  $         & 14.6  \\
			
			1 & 107.5  & $85^{+2}_{-6}$      & 6.4   & $8.5^{+0.4}_{-0.8}$       & 0.3 & $0.279\pm0.008$             & 0.51  & $1.0\pm0.1$           & 17.3  & 12.2  & $12.24^{+0.01}_{-0.51}$   & 16.2  \\
			
			2 & 27.1   & $24.2\pm0.9$        & 5.2   & $4.9\pm0.3$               & 0.3 & $0.232\pm0.007$             & 1.48  & $1.4\pm0.1$           & 17.3  & 6.6   & $6.12^{+0.51}_{-0.01}$    & 16.3  \\
		
			3 & 201.5  & $173^{+6}_{-7}$     & 10.5  & $10.52^{+0.09}_{-0.14}$   & 0.4 & $0.392^{+0.008}_{-0.006}$   & 1.01  & $1.14^{+0.06}_{-0.05}$& 14.5  & 8.2   & $7.7\pm0.5$               & 13.5  \\
			
			4 & 62.7   & $44^{+2}_{-3}$      & 5     & $6.9^{+0.4}_{-0.2}$       & 0.3 & $0.332^{+0.007}_{-0.016}$   & 1.06  & $2.1\pm0.2$           & 13    & 7.7   & $8.16^{+0.01}_{-0.51}$    & 12    \\
			
			5 & 33.0   & $13.0^{+0.3}_{-0.4}$& 1.1   & $3.6^{+0.2}_{-0.3}$       & 0.2 & $0.212\pm0.008$             & 0.16  & $1.6\pm0.2$           & 13.2  & 8.7   & $7.0\pm1.0$               & 10.5  \\
			
		\end{tabular}
	\end{center}
\end{table*}

We develop a procedure to pair groups formed in the \starforge\
simulations with associations observed by Gaia, where we remove stars that would
not be observable or that fail to meet quality indicators, such as the
renormalised unit weight error (RUWE).

For a target Gaia association and a given \starforge\ group, we follow these steps:
\begin{enumerate}
        \item Identify the model group at a point when the median age of the member
                stars is equivalent to the estimated age of the target association.
        \item Use the PARSEC isochrones to determine the approximate mass range
                corresponding to the luminosity range that is observable at the
                target cluster distance. These limits correspond to the levels
                that exceed Gaia's sensitivity and saturation limits, i.e.\
                magnitudes between 3 and 20.7 \citep{GaiaDR2}.
        \item  Remove stars from the model groups that fall outside the observable range.
        \item  Establish a threshold of 1$\arcsec$, below which we consider
                stars to be unresolved binaries or multiples. We assume that such stars
                would exhibit a high RUWE and, consequently, both members be excluded
                from the kinematic measurements.
        \item  Exclude any pair of stars in the model group that are closer than 10,000\,AU.
        \item  Account for the 15\% source loss rate measured by
                \cite{Kerr2021}, which pertains to member stars excluded due to
                quality cuts. Randomly remove 15\% of the model group members and
                repeat this process to create a sample of 100 synthetic groups for
                each model group.
\end{enumerate}

Using this method, we obtain 100 samples of each \starforge\ group. We calculate the
25th, 50th, and 75th percentiles for each measured property, which we report as
error bars. We avoid velocity contamination from binary members by excluding
unresolved and resolved binaries from the velocity dispersion and dynamical
traceback measurements \citep{Kerr2022b}. However, binaries are included in the
group's size and mass measurements. We provide a summary table with the real and sampled values, including traceback ages, in Table~\ref{tab:data} measured at 25\,Myr of simulation time.

Missing stellar mass is a known observational bias. Following \citep{Kerr2022b}, we
include a correction factor that accounts for unseen mass at the low-mass tail
of the mass distribution. We obtain the factor by integrating \cite{Chabrier2005}
in the uncovered range. For most cases in this work, which compares to observations of CFN, it
corresponds to the range between 0.01-0.09 \,\msun, which gives a correction factor
of $\sim2\%$. We also introduce a factor to account for the 15\% source loss we
remove randomly, considering an average stellar mass of $\langle m_\star\rangle =
0.5\,\msun$, i.e., correcting by a factor of 0.15$\langle m_\star\rangle$. We treat
unresolved binaries as single combined systems, only removing them either randomly
or if one of their members is not in the observable range. We
treat the members of wide binaries independently. We do not apply corrections for the binaries'
masses and only remove them following the procedure we describe above.

We compare the observed CFN systems with the \starforge\ systems by placing
the modelled systems at the same distance as the observed group. In order to
determine whether two stellar groups are equivalent, we define a parametric distance
based on mass, size, and velocity dispersion. We define the relative distance between
each observable as:

\begin{eqnarray}
        d(X) = \frac{X_{\rm obs} - X_{\rm model}}{X_{\rm obs}}, 
\end{eqnarray}
where $X$ represents the mass $\mtot$, half mass radius (\rhalf) and
one-dimensional velocity dispersion (\sigmaoned) of the target group.
Then, the normalized distance between a given observed and modelled group is 
\begin{eqnarray}
        \score  = \sqrt{ d(\mtot)^2 + d(\rhalf)^2 + d(\sigmaoned)^2 }.
\end{eqnarray}
We consider that two groups match when $\score \leq 0.6$.

\section{Results}

\subsection{Parameter measurements and their observational biases} 
\label{sec:bias}

In this section, we examine how basic quantities, which are necessary for the
characterization of a stellar group, are affected by observational biases. The
parameters we explore are the group total mass (\mtot), half-mass radius
(\rhalf), and one-dimensional velocity dispersion (\sigmaoned). Each of these
three parameters is crucial to assess the current dynamical state of stellar
groups, which can be characterized by the virial parameter. Assuming a uniform
distribution of stars, the virial parameter is given by:
\begin{equation} 
\alphavir = \frac{5\rhalf\sigmaoned^2}{G\mtot}, 
\end{equation}
where $G$ is the gravitational constant. We analyse the variation of each of
these components independently and evaluate how their errors combine to affect
the inferred dynamical state of stellar groups.

\begin{figure*}
\begin{tabular}{cc}
        \includegraphics[width=\columnwidth]{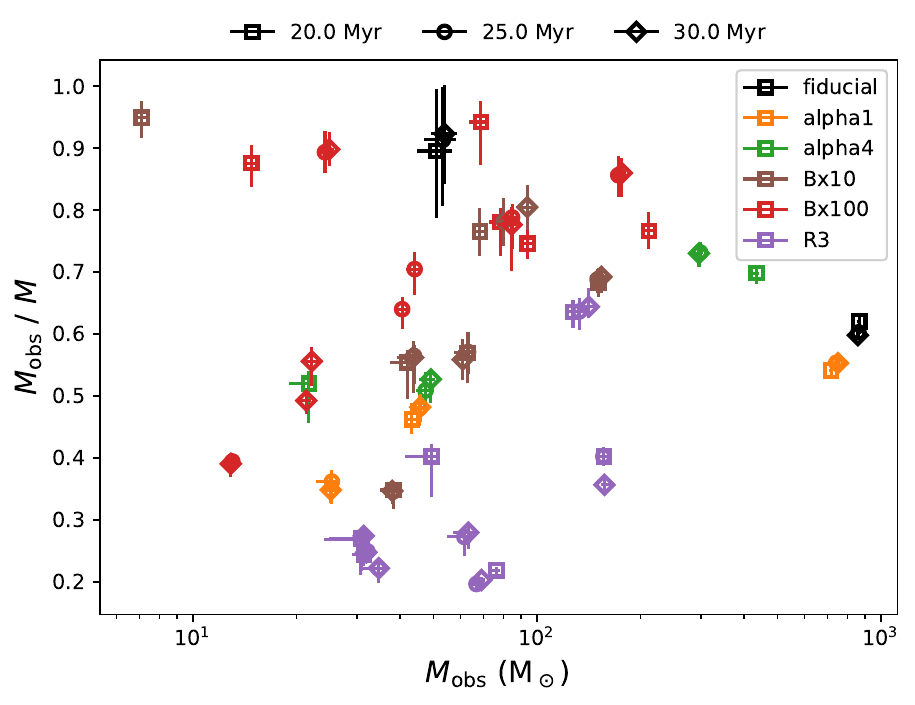} & 
        \includegraphics[width=\columnwidth]{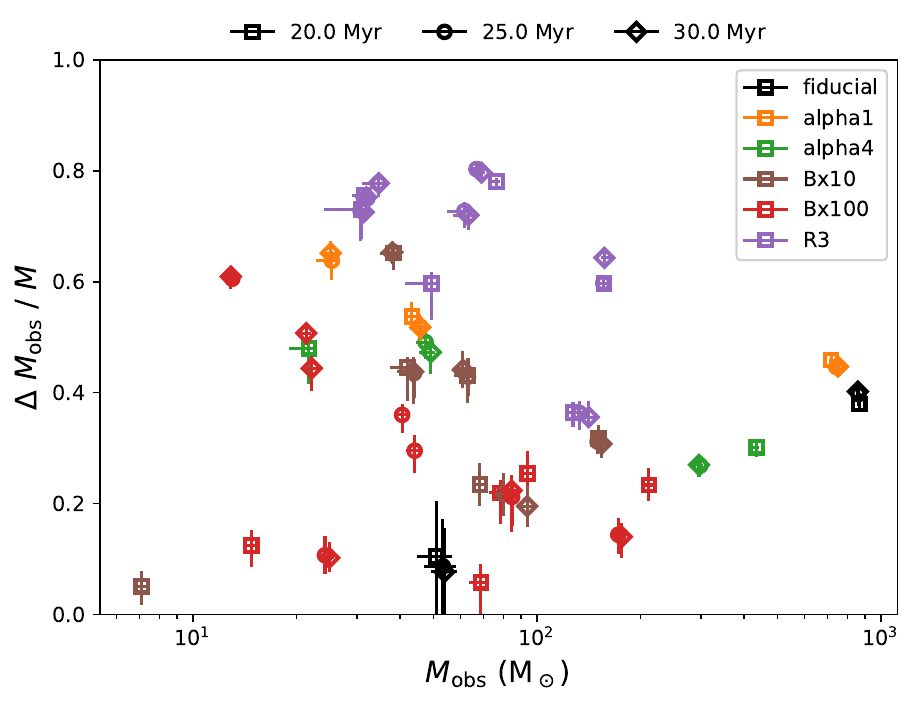} \\
        \includegraphics[width=\columnwidth]{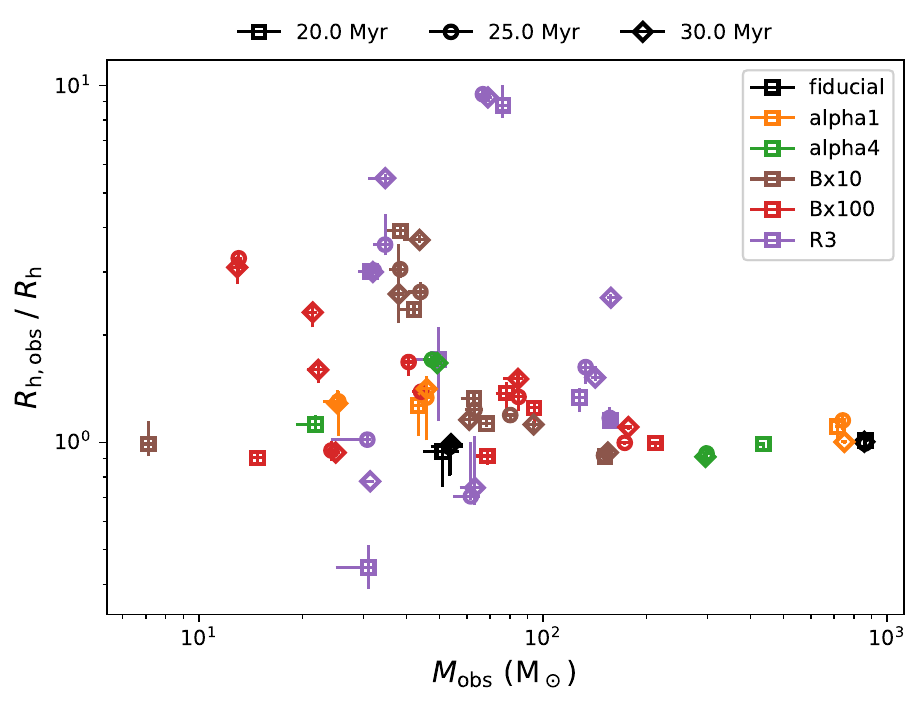} & 
        \includegraphics[width=\columnwidth]{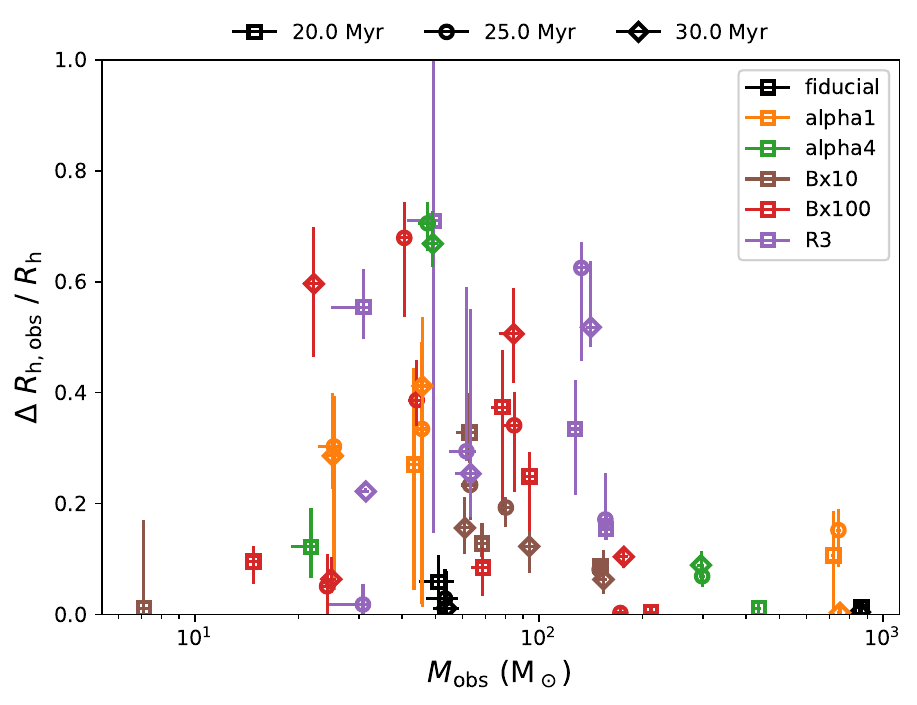}  \\
        \includegraphics[width=\columnwidth]{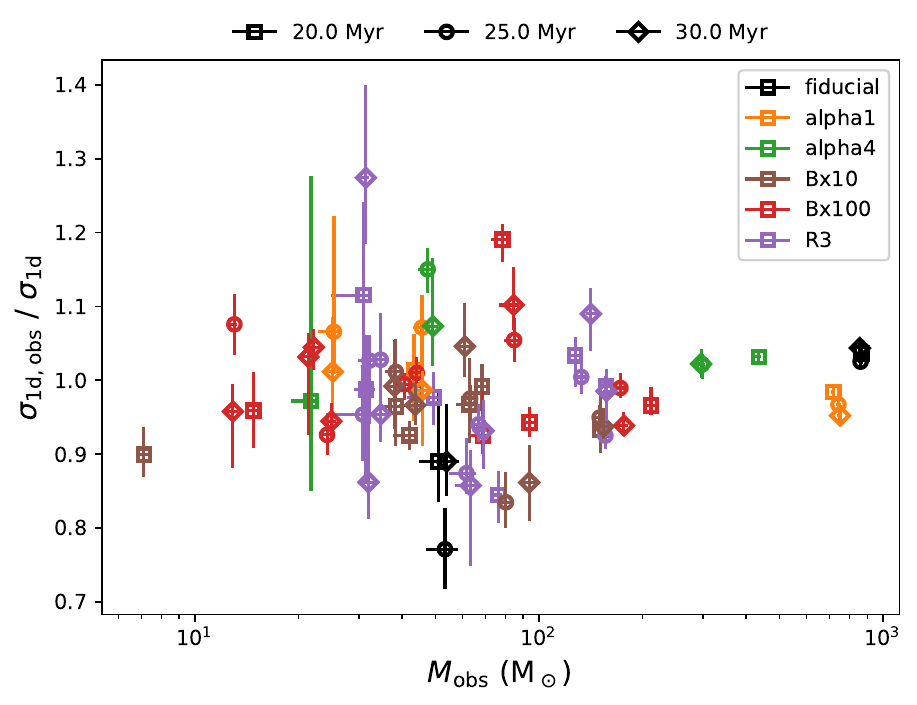} & 
        \includegraphics[width=\columnwidth]{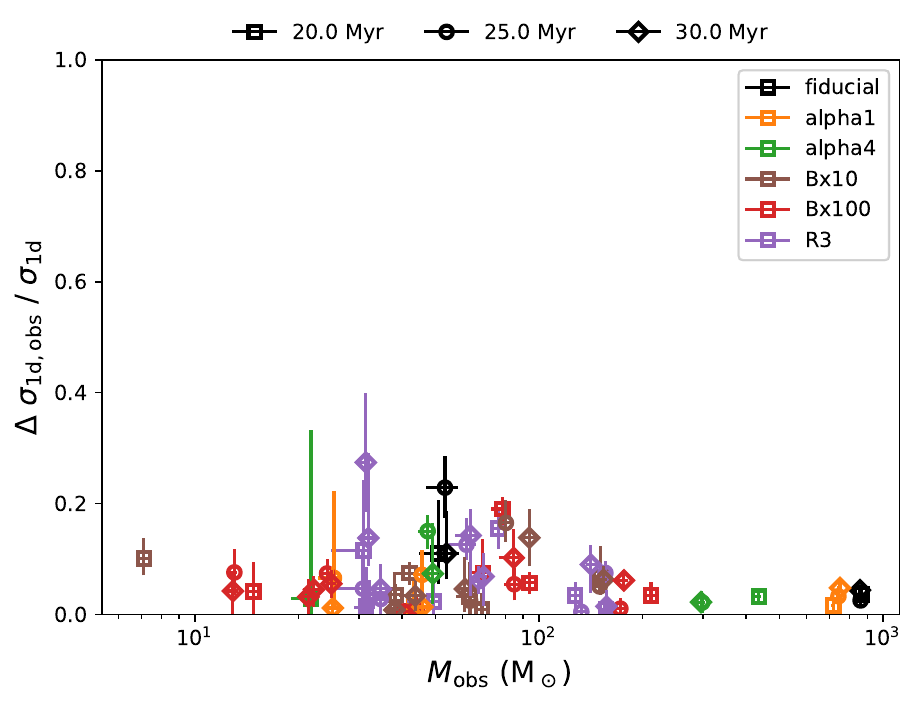} 
\end{tabular}
        \caption{ 
                Deviation of observable quantities when observational biases are
                applied as a function of observable mass for groups identified in
                \starforge. From top to bottom, total mass, half mass radius and
                velocity dispersion respectively. Left column shows deviation as a
                factor while right panel shows the deviation as relative error.
                Biasses are applied considering each system is 200 pc from the Sun.
                In general, errors decrease as systems are bigger with the total
                mass of the cluster as the most uncertain quantity with an average
                of 40\% error and the velocity dispersion the most stable with
                relative errors below 20\%.
        }
        \label{fig:errors}
\end{figure*}

Figure \ref{fig:errors} shows how these three parameters deviate from the true
values for an assumed group distance of 200\,pc. For each data-point, we repeat
the process described in \S\ref{sec:matchingprocedure} 100 times, where the
errorbars indicate the variation between the 25th and the 75th percentile. We
plot the model groups at ages of 20, 25, and 30 Myr. 
The left panels in Figure~\ref{fig:errors} show the spread in errors as an
inverse correction factor that applies to each individual observation. Here, we
see the preferential direction of the biases; for instance, masses are always
underestimated, while measurements of \rhalf are generally larger, up to a
factor of ten, than the true values. In contrast \sigmaoned errors have no
preferential direction. We find no major differences in the measured properties
at these different times. The right panels in Figure~\ref{fig:errors} show the
relative magnitude of the uncertainties. The errors in the masses, \rhalf, and
\sigmaoned, can reach 80\%, 100\%, and 40\%, respectively.

The top row in Figure~\ref{fig:errors} illustrates that the recovered mass
varies from 20\% to 90\% of the actual group mass, with an average relative
error of 40\%. Groups forming from the densest clouds (\rsmall model) have the
lowest mass recoveries. These larger errors arise because the groups contain
more massive stars, which are excluded from the calculations because their
luminosities exceed the Gaia saturation limit.

\begin{figure}
        \includegraphics[width=0.45\textwidth]{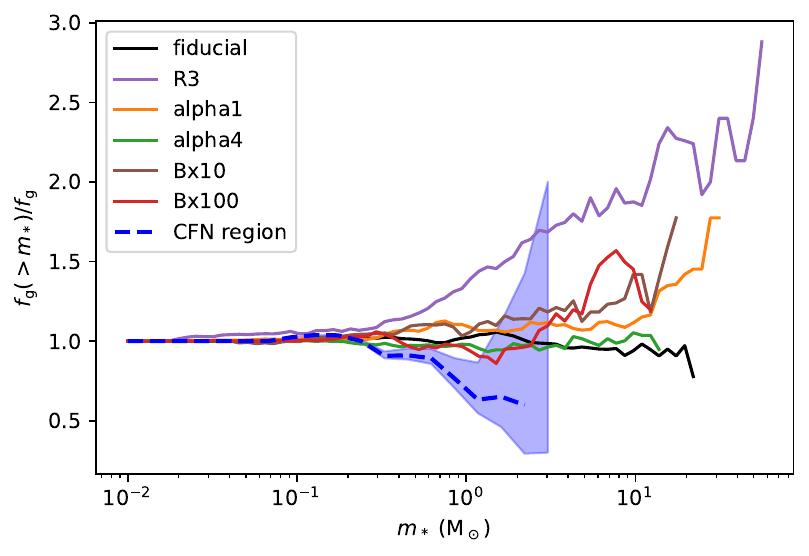 }
        \caption{
        The fraction of stars more massive than $m_*$ that are group members,
        $f_{\rm g}(>m_*)$, normalized by the total fraction of stars that are in
        groups in each simulation. As  the value of the threshold mass $m_*$
        increases, the number of stars included in $f_{\rm g}(>m_*)$ decreases;
        we only keep bins that contain more than 5 stars. The blue dashed line
        represents CFN using the available data (with masses below
        3.8\,\msun). To provide uncertainties associated with mass
        incompleteness, we also include stars more massive than 3.8\,\msun
        sampled from a \protect\citep{Chabrier2005} mass function (see text)
        showing the result of placing all those stars either inside or outside
        groups.
        }
        \label{fig:massivesingroups}
\end{figure}

To better understand how more massive stars influence the measured group mass,
we analyse how these stars are distributed in the simulations. We measure the
fraction of all stars that are part of a group relative to all stars in the
simulation ($f_{\rm g}$). Figure~\ref{fig:massivesingroups} shows the change in
the star mass fraction, $f_{\rm g}$, as a function of the threshold mass, i.e.,
$f_{\rm g}(>m_*)$ normalized by the global factor $f_{\rm g}$. At the lowest
mass limit, i.e., all stars are included, the value $f_{\rm g}(>m_*)/f_{\rm g}$
is always unity.

As the low-mass threshold increases, the fraction of stars that are members of a
group significantly increases in most models. In other words, more massive stars
are more likely to be found in groups. The \rsmall\ model exhibits the most
significant trend: most of the more massive stars are preferentially inside one
of the identified groups. We also see a similar, but weaker, trend for the
\alphaL, \Bten, and \Bhundred models. In contrast, the \fiducial\ and \alphaH
cases show no preference for more massive stars to belong to groups at any mass
range.

The fact that initially sub-virial clouds and high-density clouds show a strong
increasing trend suggests that the compositions of their groups are influenced
by dynamical encounters, i.e., as massive stars interact with other lower mass
stars, they lose energy and sink to the centre of their respective groups, an
effect known as mass segregation \citep[see][for a detailed analysis of massive
segregation in \starforge]{Guszejnov2022}. If mass segregation were primordial,
with massive stars forming in the centres of the densest areas, we would not
expect a substantial difference between the models. However, other factors may
also play a role, since models with higher magnetic fields also show a slight
preference for more massive stars to belong to groups. 

We also use the available observational data to include a line for CFN in
Figure~\ref{fig:massivesingroups}. In contrast to the simulation, the fraction
of stars in groups {\it decreases} with an increasing stellar mass threshold:
more massive stars are {\it less} likely to belong to groups. This could be
explained if more massive stars are preferentially dynamically ejected or more
likely to form in isolation. The former explanation is difficult to reconcile
with the low group stellar densities, while the latter is counter to
observational studies measuring the stellar mass function of small groups
\citep[e.g.,][]{Kirk2011,Kirk2012}. However, it is possible that observational
biases obscure the true trend with mass. We supplemented the CFN member
list with stars exceeding 3.8\,\msun, the estimated maximum resolvable stellar
mass at the CFN distance. Using a scaled \cite{Chabrier2005} mass
function—adjusted so that the total mass within the 0.09 to 3.8\,\msun\ range
matches that of CFN— we generated 100 independent mass samples. On average, each
sample includes 12 stars above this range. To evaluate how these additional
stars affect the CFN curve in Figure~\ref{fig:massivesingroups}, we consider
two extreme scenarios for each sample: either all or none of these stars were
assigned to groups. The shaded region in the figure represents the variation
between the 25th and 75th percentiles across all samples. 
In this case, the CFN curve is marginally consistent with unity, i.e., no mass
dependence on group membership. In contrast to a dynamically rich
environment, such as the \rsmall model, the tendency of massive stars to
preferentially reside in groups appears at relatively low masses, e.g. at 1
\msun. However, at this mass threshold the CFN region shows the  inverse trend
-- even in the extreme test case where all the (hypothetical missing) massive
stars are placed in groups (see the blue-shaded region in Figure
\ref{fig:massivesingroups}). This result suggests that the CFN complex either
did not go through a high density phase or that massive stars simply did not
form in the region.
A more detailed study is needed to verify the magnitude of mass incompleteness
on groups in CFN.

In contrast, the measured group half-mass radius is less clearly impacted by the
completeness of high-mass group members, but more prone to systematic
uncertainties. The middle panels of Figure \ref{fig:errors} show the errors in
$R_{\rm h}$ as a function of the inferred group mass. For groups with masses
below 100\,\msun, the observed \rhalf\ can be overestimated by up to a
factor of 10. This error decreases as groups grow in mass, and therefore in
number, with 100~\msun being the threshold at which \rhalf\ becomes reliable (in
our sample). The relative errors for groups below 100~\msun\ spread up to 80\%,
while the uncertainty in radius for larger groups is generally below 20\%

The bottom panels of Figure~\ref{fig:errors} show the error variation in
\sigmaoned. The accuracy of the observed \sigmaoned also depends on the group
mass. However, the measured \sigmaoned tends to be more accurate across all
group masses than either \rhalf\ or \mtot. The \sigmaoned correction factor lies
between 0.7 to 1.3 and approaches unity as \mtot\ increases. The relative errors
are below $\sim$20\% for all groups and less than $\sim$5\% for the most massive
groups. Consequently, \sigmaoned\ is relatively unaffected by observational
biases and is the most reliable parameter for observed stellar associations.

\begin{figure*}
        \includegraphics[width=\textwidth]{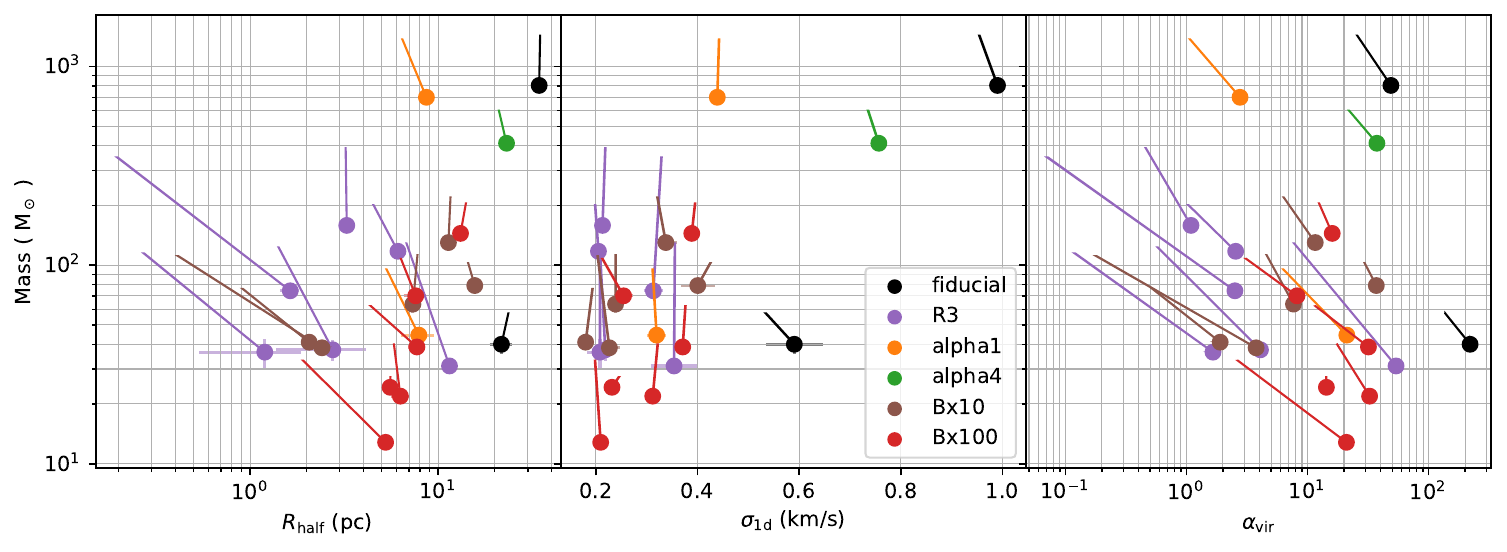}
        \caption{
        Properties of the groups identified in the \starforge\ simulations and
        their parameter shift after applying observational biases. Stellar
        groups are selected at an age of 20\,Myr and placed at a distance of 200
        pc, a similar distance to and age of groups in the Cepheus Far North
        complex. Filled circles show the value of each parameter after the
        sampling process, while lines connect to their true values. Order of
        magnitude errors in radius, mass, and virial parameter are common for
        small groups.
        }
        \label{fig:allmodels}
\end{figure*}

Figure~\ref{fig:allmodels} shows how the errors in the three observed cluster
parameters correlate with one another. The lines in Figure~\ref{fig:allmodels}
show the movement of groups in the mass-size and mass-\sigmaoned\ diagrams after
applying observational biases. We display all groups from all \starforge\ models
at an age of 20\,Myr and assume a distance of 200\,pc. While the most
significant change occurs in the derived mass, the left panel shows a systematic
trend for low-mass clusters to have systematically over-estimated radii. This is
because their radial distribution is intrinsically poorly sampled. 

The combination of these effects produces a large systematic increase in the
estimated virial ratio. As shown in the third panel of Figure
\ref{fig:allmodels}, errors in the underlying parameters can produce an observed
virial ratio of low-mass clusters that is significantly overestimated; i.e.,
small stellar groups are likely much more bound than they seem based on their
observed masses, radii, and velocity dispersions. Assuming the error in
\sigmaoned\ is small, the radius is overestimated by a factor of 2, and only
60\% of the actual group mass is recovered results in an observed \alphavir that
is 3.3 times larger than the true value.

\begin{figure}
        \includegraphics[height=2.7in]{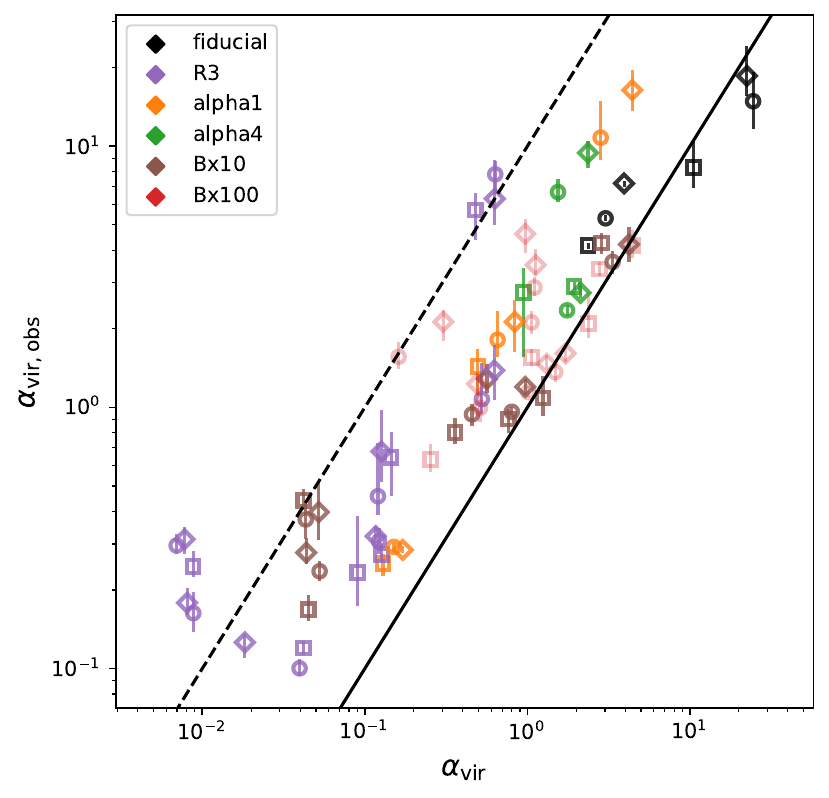}
        \caption{
        One-to-one comparison between the virial parameter calculated using all
        stars ($\alpha_{\rm vir}$) and the equivalent using only
        observable stars ($\alpha_{\rm vir, obs}$), with the solid line representing equality.
        Errorbars represent the range
        between 25-75 percentile caused by the random sampling of unseen stars for
        groups in all simulations measured at 20, 25, and 30 Myr.
        After introducing observational biases $\alpha_{\rm vir,obs}$
        is up to 10 times (dashed line) higher than $\alpha_{\rm vir}$. 
        }
        \label{fig:virialratio}

\end{figure}

Figure~\ref{fig:virialratio} shows a direct comparison between the virial ratio
obtained when using all stars ($\alpha_{\rm vir}$) and the virial ratio derived
after introducing observational biases ($\alpha_{\rm vir obs}$). We show the
same groups as in Figure~\ref{fig:errors}, i.e., groups measured at 20, 25, and
30 Myr. In general, the virial ratio is overestimated by a factor of ten in our
sample. This result suggests that the virial ratio of observed stellar systems
is significantly overestimated.

\subsection{Cepheus Far North group analogues }
\label{sec:cfn}

\begin{figure*}
        \includegraphics[width=\textwidth]{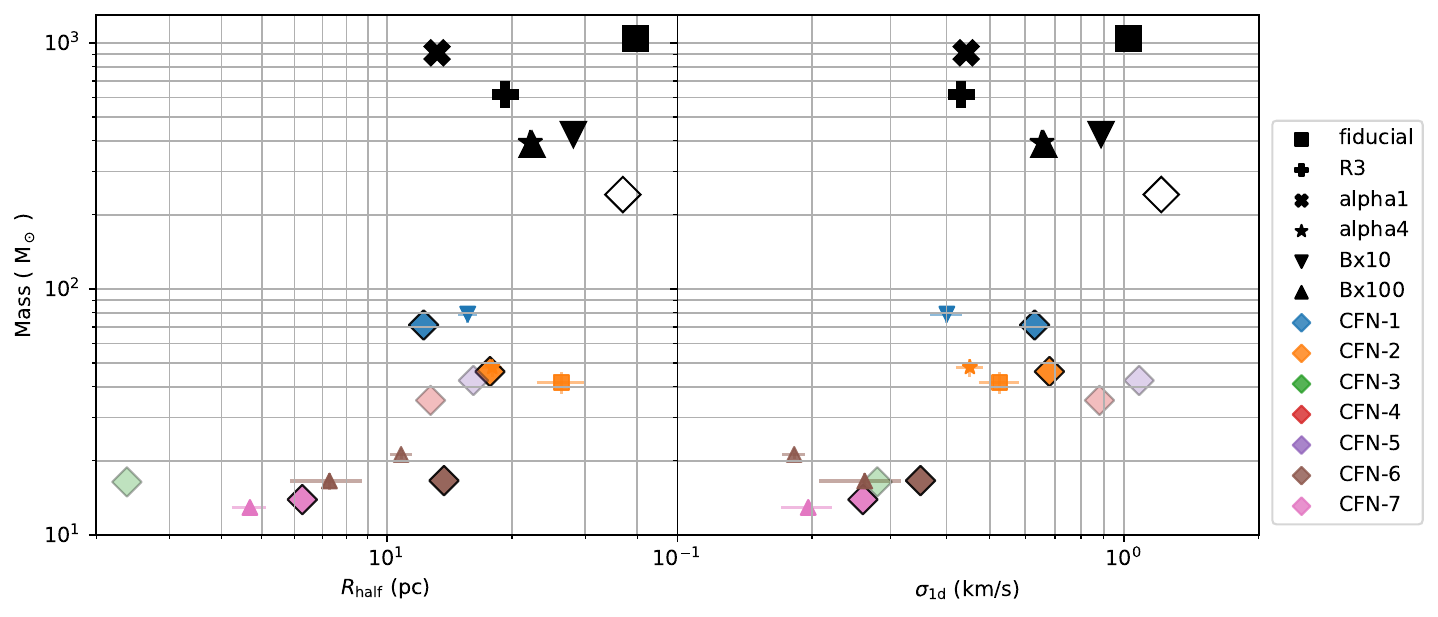}
\caption{Best group-matched models. Diamond symbols represent the target CFN
groups. Symbols of the same colour indicate model groups that match CFN groups
when their ages are equivalent and placed at the same distance from the Sun. The
symbols indicate the model from which the group originates, as shown in the
legend. Black symbols represent the properties of the entire CFN region and
\starforge\ models, considering all observable stars.}
        \label{fig:matches}
\end{figure*}

As shown in previous works, the \starforge\ simulations assuming typical galactic
cloud conditions tend to form large, expanding associations.  The CFN
association appears to be a close analogue to the simulation results.
Figure~\ref{fig:matches} shows the global parameters of CFN (represented by the
empty large diamond) compared to those of the stellar complexes formed in the
\starforge\ simulations (black filled symbols). In all cases, we match the median
ages of the stars to the median age of the groups in CFN. We only consider stars
that are part of a group for calculating these parameters.

\begin{figure*}
\includegraphics[width=\textwidth]{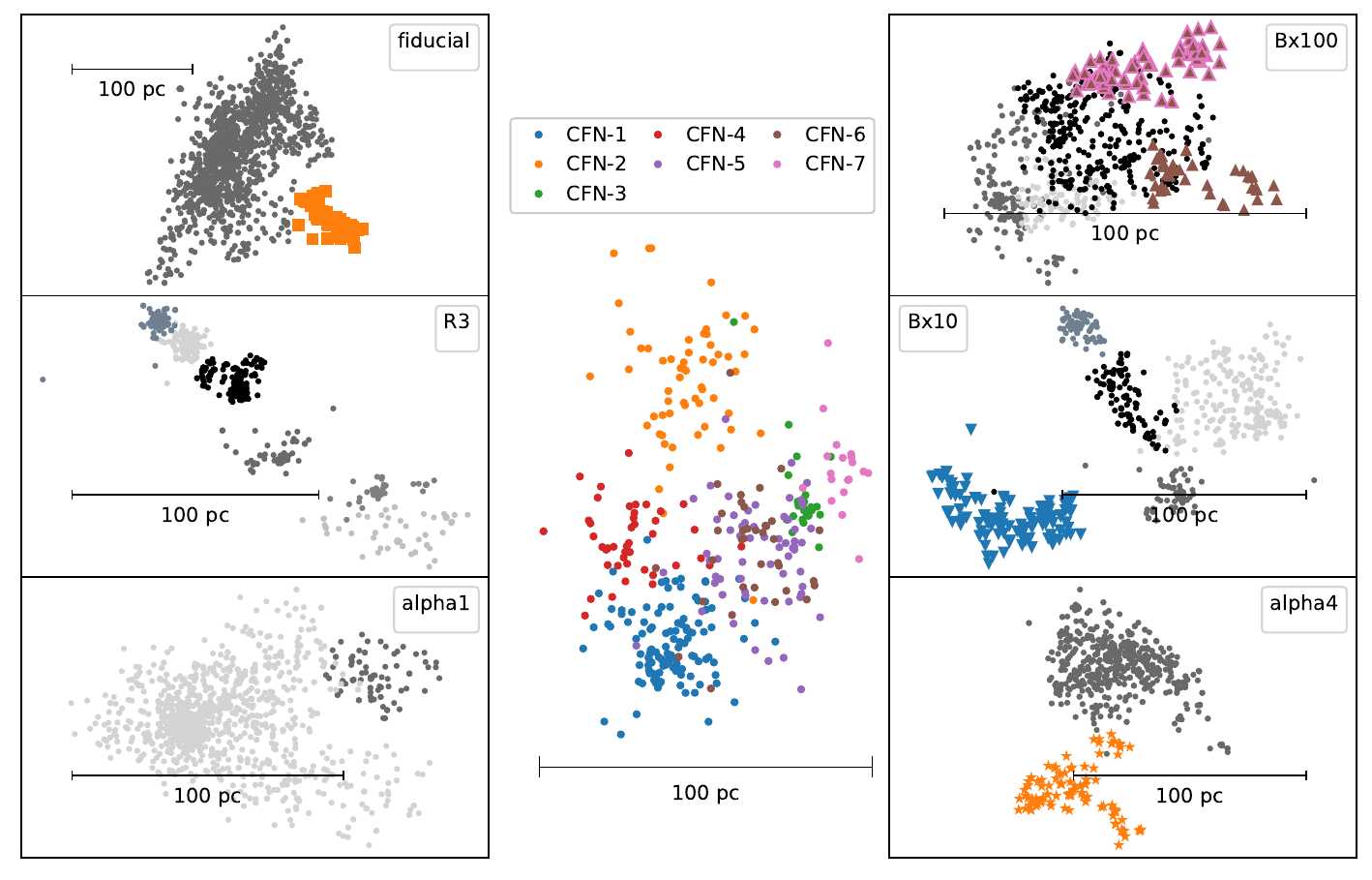} 
\caption{Cepheus Far North groups as labelled in \protect\cite{Kerr2022b} (centre).
Side panels shows the distribution of groups from different models after
applying observational biases and at the time where the median age of the stars
in each model is 25 Myr (\rsmall\ and \alphaL). Clusters that match our criteria
have the same colour as the groups of CFN shown in the centre. Note that one of
the groups in the \Bhundred\ model matches CFN-6 and CFN-7 simultaneously but at
at 24 and 17.1 Myr respectively. 
}
\label{fig:cfnoverview}
\end{figure*}

The CFN complex has a global half-mass radius of about 36\,pc with a total mass of
$\sim250$\,\msun\ spread across seven groups. The \starforge\ associations analyzed
here have half-mass radii ranging between 10 and 40\,pc, forming between 3 and 8
groups with total masses between 300 and 1000\,\msun. While the $\rhalf$ comparison
should be taken with care, as it is highly dependent on each region's internal
substructure, Figure~\ref{fig:cfnoverview} shows the spatial distribution of the
modelled regions and the CFN. Visual inspection suggests that, despite the different
values of $\rhalf$, these regions span similar sizes of about 100 to 300
pc. Additionally, CFN appears significantly sparser, which accounts for
its large half-mass radius.

CFN subgroups one-dimensional velocity dispersions range between 0.26 and
1.08\,\kms, which is within the same range as groups identified in the
\starforge\ simulations, with an average \sigmaoned of 0.5\,\kms
\citep{Farias2023a}. However, the CFN region's global velocity dispersion is
somewhat higher than the global regions obtained from the simulations, at
$1.2\pm\mathbf{0.2}$\,\kms when calculated using transverse de-projected velocities. From the
position of CFN in the parameter space shown in Figure~\ref{fig:matches} we see
that CFN is slightly sparser and more loosely bound than complexes formed in the
simulations, while at the same time, CFN groups appear in the same parameter
space region than groups formed in \starforge. This suggests that while the
processes that formed the subgroups may be similar to \starforge, the CFN natal
cloud likely had either higher virial ratio, slightly lower global surface
density and/or a lower star formation efficiency.

Following the matching process described in \S~\ref{sec:matching}, we find that four
out of the seven CFN groups have good matches to groups within ${\cal D}\leq 0.6$ in the
\starforge\ simulations. Figure~\ref{fig:matches} shows the position in the mass-size
and mass-velocity dispersion diagrams for all the matching groups. For reference,
we also show CFN groups with no match as semi-transparent symbols.

While, in general, the modelled stellar complexes are similar to CFN, there is
no one simulation that produces groups that match most CFN groups. The most
successful model is \Bhundred, i.e.\ increasing the fiducial magnetic field a
factor of ten, which forms three groups that match CFN associations, although
two of these groups match the same CFN group (CFN-6) but at different epochs. As
shown in our previous study, the \Bhundred\ model tends to form a few small
bound clusters on the order of 20 to 60\,\msun, which are considerably smaller
than the one or two clusters formed in the fiducial cases with masses of
$\sim800$\,\msun. As explained in \cite{Guszejnov2022a}, the cloud in the
strongest magnetic field simulation (\Bhundred) is not unbound by stellar
feedback and protostellar jets. It only begins to disperse after the first
supernova explosion. This results in a lower star formation rate but also causes
stars to form at later stages when the cloud is still bound but more diffuse.
Stars formed at these later times are assigned to groups that remain
well-separated and do not have time to merge with other groups before cloud
dispersal, yielding to low-mass clusters in an expanding region.

CFN groups that do not match any of the modelled clusters have either a
significantly smaller half-mass radius (CFN-3) or a high velocity dispersion
(CFN-5 and CFN-7). Modelled groups typically have velocity dispersions below
0.5\,\kms, while CFN-5 and CFN-7 have dispersions above 0.7\,\kms. The former
have velocity dispersions of the same order as the global region, This means
that those stars may not be independent groups but rather stars that are simply
expanding along with the region.. However, the masses of these groups are in the
same range as those of other modelled groups.

The formation scenario for groups in CFN is consistent with the formation
mechanism presented by the \starforge\ simulations. While the match is not exact,
our analysis suggests that the conditions of the cloud that gave birth to CFN
are similar to those of the \starforge\ clouds, i.e., of a typical Galactic cloud,
which has properties similar to present-day nearby star-forming regions such as
the Perseus and Orion molecular clouds.

\subsection{Traceback Ages}

In recent years, there has been increased study and discovery of expanding
associations, given the unprecedented kinematic accuracy provided by Gaia. This
has motivated analyses that ``trace back" the trajectories of the association
members to identify the time at which the stars were most compact, i.e., an
indication of their formation time and therefore their age
\citep{Miret-Roig2020,Galli2023}. traceback studies provide a photometrically
independent estimate of association ages. In general, these derived dynamical
ages are slightly smaller than photometric ages, because the traceback age
estimates the time the stars begin to expand. By definition, most of the
association members are already formed, since expansion occurs during the
dispersal of the natal cloud via stellar feedback or supernova explosions
\cite{Miret-Roig2024}.

However, estimating the traceback time requires a number of assumptions. For
instance, the traceback of trajectories is usually done while ignoring the
interactions between stars, although most studies do take into account the
gravitational potential of the Galaxy. Such an approach is only valid for
expanding associations that are completely or mostly unbound. Estimates also
assume that the expansion rate of the systems is constant from their origin
\citep{Crundall2019}. While these are reasonable assumptions, it is not clear
how much associations deviate from these and how this may affect the estimated
ages.

Given the complete history of the modelled expanding stellar complexes provided by
\starforge, we investigate the relationship between the observational traceback age
(\tdyn) and the true traceback age (\tdyntrue) as described in
\S\ref{sec:agemethods}. 
To obtain \tdyntrue we do not make any assumptions, as it follows the true trajectories of
their members in the simulation, i.e.\ the fraction of bound stars and any change in the
expansion rate of the region, do not affect the result.

\begin{figure}
        \includegraphics[width=\columnwidth]{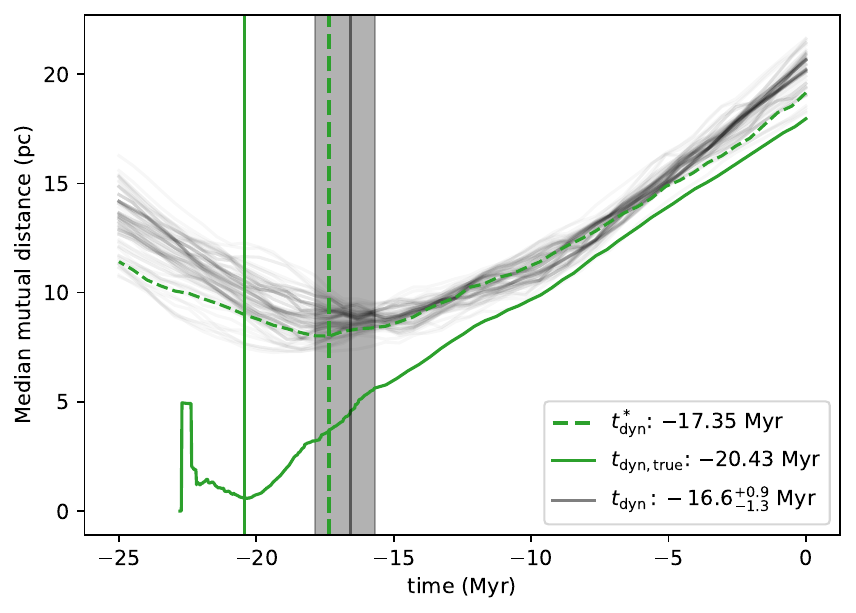}
        \caption{
        Median mutual distance versus time inferred from tracing back a typical
        stellar association. The solid line represents the true history based on
        the simulation data. Dashed lines indicate the traceback estimations
        using all available stars, excluding binary systems. Semi-transparent
        lines show the resulting histories after applying observational biases
        100 times. Vertical green solid and dashed lines mark the locations of
        the minimum median mutual distance for the first two cases. The shaded
        area represents space between the 25th and 75th percentiles of the
        sampled case with the median shown in a black vertical line.}

        \label{fig:tbexample}
\end{figure}

Figure~\ref{fig:tbexample} shows the median mutual distance history of a group
from the \alphaL\ model compared to the estimated expansion based on their
velocities at 25\,Myr. The traceback analysis suggests that the group began its
expansion at approximately \textbf{-17.35}\,Myr, with a median mutual distance
of 7.5\,pc. However, the actual simulation data shows that the group begins
expanding at \textbf{-20.4}\,Myr, when the median mutual distance was less than 1\,pc. The
figure shows that the trajectories significantly diverge after about 10\,Myr of
trajectory traceback. The difference occurs because the traceback estimate does
not account for the stellar relaxation that occurs due to the influence of the
dispersing gas. The trajectories of the stars during the early expansion
\textbf{(-21 to -15\,Myr)} are affected by both the residual gas and also by the forces
between stars, which, even though unbound, still exert influence. Given the
chaotic nature of the $N$-body interactions and perturbations from the gas it is
not possible to tightly constrain the traceback trajectories. The grey lines in
Figure~\ref{fig:tbexample} show the effect of re-sampling the stars 100 times,
to account for the 15\% stellar completeness described in \S~\ref{sec:matching}.
In this example, excluding 15\% of the association members, in addition to other
members that are not observable, results in a variation of the estimated
traceback time by $\sim$2.5\,Myr. This uncertainty is typically within range of
the best case scenario, which includes all group members and ignores
observational biases. This means that both $\tdyn$ and $\tdyntrue$ are
mostly affected by the weak interactions between stars and early relaxation,
with completeness playing a secondary role.

\begin{figure*}
        \includegraphics[width = 0.9\textwidth]{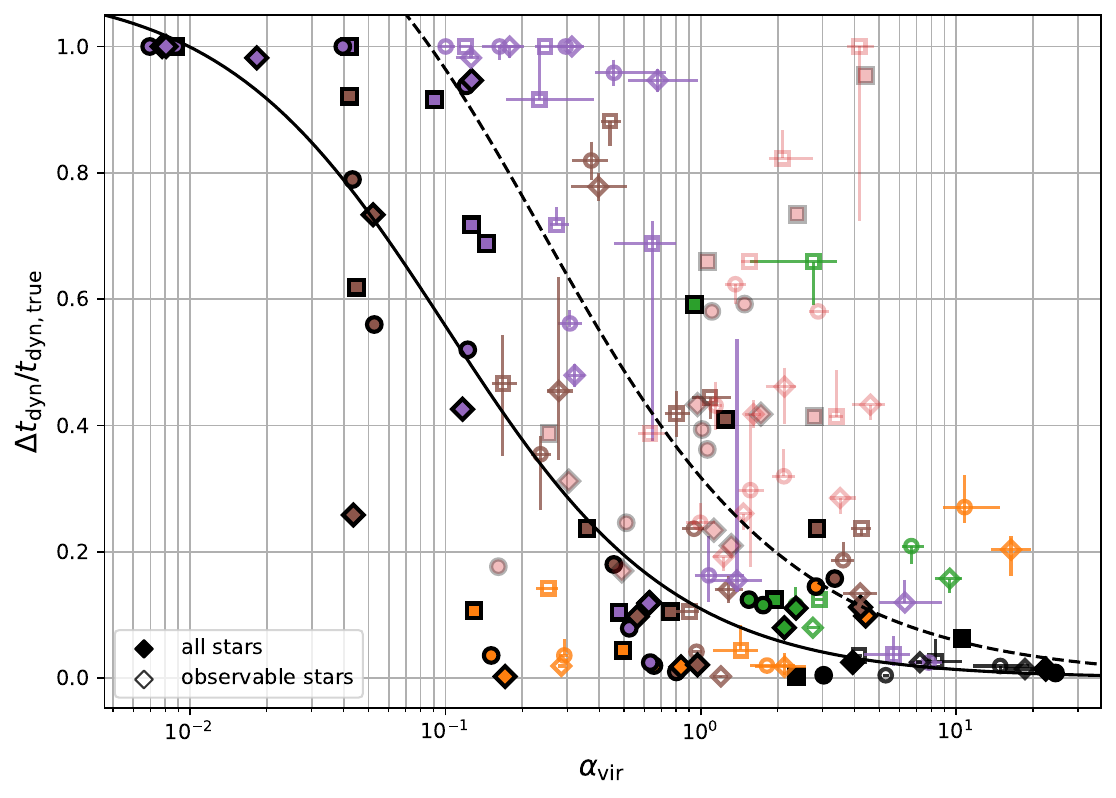}
        \caption{
                Fractional error of the traceback dynamical age
                versus group virial ratio. Filled symbols show measurements of
                the traceback age and virial ratio that include all stars. Empty
                symbols show the same measurements but using the sampling
                prescription described in \S~\ref{sec:matching}. Errorbars
                represents the range between 25-75 percentile caused by the
                random sampling. The solid line represents a softened power law
                fit for the case with all stars (excluding the \Bhundred\ case,
                see text) and a dashed line indicates a fit to  the sampled
                case. 
                We show the measurements at 20, 25 and 30 Myr from the start of the
                \starforge\ simulation using squares, circles, and diamonds,
                respectively.
        }
        \label{fig:tbackerror}
\end{figure*}

While most of the groups identified in \starforge\ are unbound, some contain
subsets of stars that are loosely bound. These members reduce the accuracy of
the inferred traceback time, since they expand more slowly or not at all. Groups
that are completely bound have a traceback age of zero. Naturally, we expect
that more unbound systems have fewer internal dynamical interactions and thus
have a more reliable \tdyn. Figure~\ref{fig:tbackerror} shows
the fractional error of the traceback time versus the virial ratio of the system
calculated utilizing all members, $\alpha_{\rm vir}$. The filled symbols
indicate the dependence as a function of the true virial parameter, i.e., where
we use all stars to compute $\alpha_{\rm vir}$; the open symbols show the result
where we apply observational biases to the calculation of $\alpha_{\rm vir}$. In
the former case, the fractional error dramatically decreases as $\alpha_{\rm
vir}$ increases, such that the error is less than 20\% for groups with virial
parameters greater than 2. However, when observational biases are taken into
consideration, the relationship between virial ratio and error shifts to higher
virial ratios and becomes less correlated; there are some groups with virial
ratios above 10 that have errors above 20\%. We  exclude the \Bhundred from
this analysis, as we find that some of the groups formed in that model contain
a different source of contamination. Star formation in the \Bhundred\ model
spans a long period of time, and a significant fraction of the stars form after
the region expands. These late-forming stars become part of the expanding groups
but their traced trajectories are not accurate and are hard to associate with
the rest of the group. Therefore, we exclude these groups from this analysis as
the traceback error of these groups is not related to the virial parameter.

\begin{figure*}
        \includegraphics[width=\textwidth]{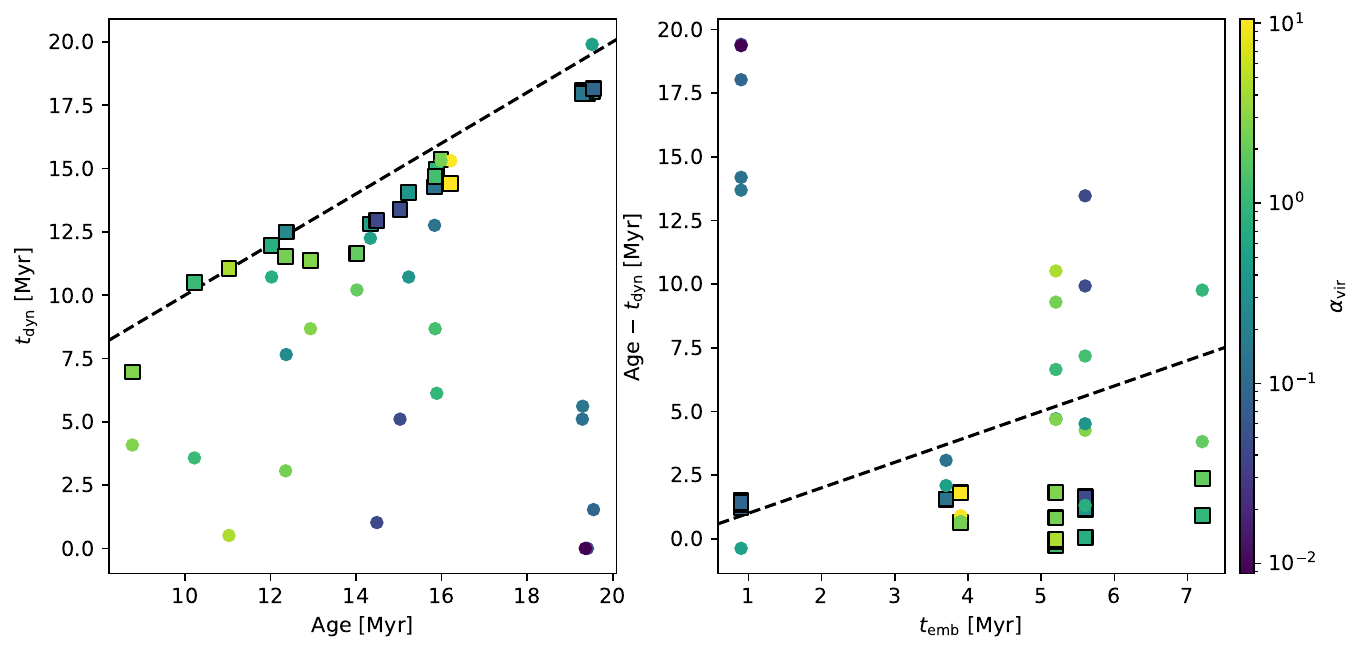}
        \caption{ \textbf{Left:} Comparison between the dynamical age ($t_{\rm
                dyn}$) and the median age of stars in each group, coloured by
        the true group virial ratio. Squares represent the \emph{true}
        dynamical age, measured directly from the simulations, while circles
        represent the dynamical age obtained by tracing back the stars linearly from their status at time $t=25$\,Myr.
        \textbf{Right:} Difference between stellar age and dynamical age compared to the
        true \emph{embedded} phase of each cluster. Dashed lines show the
        one-to-one ratio for reference. We find no clear correlation between
        these quantities. }
        \label{fig:agetraceback}
\end{figure*}

The right panel of Figure~\ref{fig:agetraceback} compares the true $\alpha_{\rm
vir}$ and its observable counterpart $\alpha_{\rm vir,obs}$, indicating that the
observed virial ratios can be overestimated by up to a factor of 10. This result
suggests that the inferred virial parameters of some associations may be
significantly overestimated, which has implications for the derivation and
accuracy of the traceback age.

We expect that $\tdyn$ correlates with the age of the stellar population;
however, $\tdyn$ should be slightly shorter, as it measures the time since the
expansion of the cluster. We show this correlation in
Figure~\ref{fig:agetraceback}, where we define the age of the groups as the median
age of their members. A strong correlation exists between these two timescales
when no observational biases are considered. However, as we introduce these
biases, the correlation vanishes, with the virial ratio of the systems seeming
to play a small role. The right panel of Figure~\ref{fig:agetraceback}
illustrates the relation between the dynamical-stellar age difference and the
duration of the embedded phase ($t_{\rm emb}$). We measure $t_{\rm emb}$
directly from the simulations as the time between the formation of the first
star and the beginning of the region's expansion. We do not find any
correlation, independent of whether we introduce any observational biases. We
note that the true value of $t_{\rm emb}$ remains somewhat ill-defined, as gas
does not vanish instantly from the cluster \citep[see][]{Farias2023a}. In our
simulations, the stars begin to expand when the cluster reaches its densest
phase, at which point most stars have formed and stellar feedback rapidly
disperses the gas within the central region of the proto-cluster. All models
with $\tdyntrue$ show an age difference below 2.5 Myr, regardless of the
duration of $t_{\rm emb}$. However, $\tdyn$ shows no correlation with $t_{\rm
emb}$, regardless of the group virial ratio. The lack of relationship arises
because star formation continues after the cloud begins to expand. 


\section{Discussion}

In recent years, the number of studies targeting faint, low-density stellar
associations has increased dramatically. The kinematic accuracy of Gaia and the
advent of more reliable clustering algorithms, such as DBSCAN, have
significantly enhanced the discovery of sparse young associations \citep[see
e.g.][]{Rizzuto2011,Cantat-Gaudin2019,Krolikowski2021}. These stellar systems
may span dozens of parsecs and show a significant level of substructure,
containing several groups that may be separated by position, kinematics, or
age \citep{Wright2020}. In this and previous works, we find that
simulations of a single cloud with realistic physical mechanisms and initial
conditions that are typical of clouds in the solar neighbourhood can produce
these substructures in a single star-formation event.

To compare the structures formed in the \starforge\ simulations, we post-processed
the stellar distributions, taking into account typical observational limitations
including photometric sensitivity, unresolved binaries, and stars that would be
excluded by typical quality cuts from databases like Gaia. Other observational
biases that interfere with constructing an accurate group census include
reddening, which might place some stars outside the detection range
\citep{Rivilla2013}, and stars in highly populated areas where point source
observations are not accurately resolved \citep{Brandeker2019}. 
Furthermore, only a small fraction of Gaia sources have radial velocities,
broadly about 2\% \citep{GaiaDR2}, and those that do have 
an order of magnitude larger uncertainty of $\sim$1 \kms compared to the typical
0.1 mas yr$^{-1}$ uncertainty in proper motion (which is about 0.2 \kms for the
distances considered here). This implies that only a small fraction of sources are
accurate enough to be used for kinematical inferences such as the traceback
age. However, the effects we include here are general and provide a basis for
understanding observational biases that illustrate how a small fraction of
missing members affect the derived properties of these observed systems.

For instance, we find that the masses of the systems are severely affected by
completeness. While we recover more than 60\% of the mass of most groups, a
significant fraction of identified groups have mass recovery that falls below
50\%. We find that this effect is most important in low-mass groups that are
born in dense environments, i.e., those that have undergone more dynamical
evolution. This is because, in such small systems, massive stars play a
significant role in forming groups, as they attract other lower-mass stars and
help keep the groups together as the region expands. Since such stars are bright
and/or usually in binaries \citep{Offner2022}, they are commonly outside Gaia's
photometric range and therefore excluded, resulting in the loss of a significant
fraction of the total group mass. 

However, the role of massive stars provides insights into the dynamical history
of stellar associations. We find that massive stars act as catalysts to form
groups in regions where dynamical evolution is more likely, such as initially
dense or virialized clouds. Therefore, massive stars are more frequently found
in groups rather than in isolation compared to lower-mass stars. This is
consistent with observations of small young groups in nearby star-forming
regions, which appear to be mildly mass segregated and where higher mass stars
tend to be more clustered \citep{Kirk2011,Kirk2012}.  However, this effect
contrasts with studies showing that massive stars are also more likely to be
ejected as runaway stars from highly dynamical regions
\citep{Hoogerwerf2000,Oh2016,Fujii2022}. It is possible that the dynamical
interactions that would have ejected those massive stars were unresolved by the
softening force that \starforge\ uses at close distances. Nonetheless, the
difference could also depend on the amount of dynamical evolution. While massive
stars are frequently involved in encounters, a large number of interactions are
needed to completely eject a massive star, and this likely requires the presence
of other similar stars. Stellar associations  are unlikely to satisfy these
conditions as their dynamical evolution is mild, and they likely expand before
massive stars have a chance to be ejected \citep{Oh2016}.

In the CFN region, there are no stars more massive than 4\,\msun\  and the
higher mass members do not appear to be preferentially in groups. While the
apparent distribution of massive stars could be affected by completeness, this
result suggests that the CFN region has not undergone significant dynamical
evolution and that the CFN stars likely formed in a low-density environment.
However, if there are indeed no stars more massive than 4\,\msun\ in the region,
this further supports the idea that this region formed in a low-density
environment and/or where more massive stars simply did not form
\citep{GrudicOffner2023}. A more complete census of the region is required to
verify the completeness of the high-mass end of the stellar population.

The kinematic accuracy of Gaia has prompted a variety of studies that aim to
trace the trajectories of stars to their initial configuration. In recent work,
\cite{Miret-Roig2024}  compared the traceback ages of six young associations to
the photometric ages derived from stellar evolution models. They found an age
discrepancy of 5.5\,Myr on average, which is a factor of four larger than the
values obtained here, with an average of $\sim1.25$\,Myr in our best case
without including observational biases. Adding observational biases
significantly increases the dispersion in predicted traceback ages. While some
additional differences may be due to the assumption that photometric ages and
simulation stellar ages are the same, we do have differences in the
interpretation. \cite{Miret-Roig2024} suggested that the photometric-traceback
age discrepancy is related to the duration of the embedded phase of
associations, i.e., the star formation phase where natal gas dominates the
potential of the star-forming region. However, we do not find evidence of this
in the \starforge\ simulations. In all our models, we find that the difference
between the dynamical and true ages is remarkably similar, despite the wide
duration of the embedded phase in the different models. We argue that the
discrepancy between ages represents the difference between the median stellar
age and the time since the stars began their expansion. While several previous
studies have assumed that gas expulsion happens rapidly and by the end of the
star formation process
\citep[e.g.][]{Baumgardt2007,Moeckel2010,Pfalzner2013,Farias2018b}, there is
growing observational evidence that stellar associations show non-isotropic
expansion \citep{Cantat-Gaudin2019,Wright2019a}, probably produced by
hierarchical star formation and/or low-density and already expanding
environments \citep{Gouliermis2018,Ward2020}. \cite{Farias2023a} show that in
the \starforge\ simulations gas removal from the stellar systems begins early in
the formation process, before stars reach their most compact configuration, and
star formation continues after the stars begin to expand. The complexity of this
process implies that there is no simple relationship between the embedded phase
length and the age discrepancy between traceback and photometric age estimates.

Despite this complexity, there is one major factor that influences the accuracy
of the traceback age estimation: the dynamical state of the expanding region. We
have shown that when the viral parameter is higher than $\approx2$ the
uncertainties in the traceback ages are typically less than 20\%. However, this
confidence requires an accurate measurement of the virial parameter, which can
be in itself challenging. While some studies have focused on characterizing the
dynamical state of stellar clusters and associations
\citep{DaRioTan2017,Kuhn2019,Wright2024}, studies utilizing the traceback-age
method often assume the systems are unbound \citep{Crundall2019,Miret-Roig2022}.
However, our estimates of errors in the group size, mass and velocity dispersion
indicate that an observed high virial ratio ($\alpha_{\rm vir} > 2$) could in
fact be an artifact of observational biases, i.e. observed associations that
appear unbound could actually have a significant fraction of loosely bound stars
that impact the estimated traceback age. This is especially important for
younger stellar associations that may contain residual gas from the parent cloud
as noted by \cite{Couture2023}. 

Some of this uncertainty could be mitigated
with more robust forms of the traceback age method. Recently,
\cite{pelkonen_evaporation_2024} proposed a new version of this method, in which
individual stars are associated with a traceback age. Using their method, it is
possible to remove outliers from the sample, thereby producing a more robust
traceback age. While here we  focus on the general traceback method, we apply
their approach in a few cases and find that it improves the accuracy of the
traceback age in cases where there are enough stars with valid trajectories—i.e.
in the larger clusters of our sample and where the virial parameter is high.
However,  it appears less reliable in the case of smaller clusters. Additional
work is required to test this method more thoroughly in the future. 

Most of the uncertainty in the virial parameter comes from the missing
massive stars due to the Gaia saturation limit. Considering a \cite{Chabrier2005}
initial mass function and a 3.8\,\msun limit, we expect about 3\% of the total
mass of a group to lie above this limit, which would imply that a group is $\sim3$\%
more strongly bound (assuming the radius and velocity dispersion do not change).
However, if massive stars are found more commonly within groups in a region,
these figures could be larger. Therefore, in order to assess the reliability of
traceback ages, it becomes important to complement membership lists with other
catalogues, such as HIPPARCOS, even if the kinematics cannot be improved.

Consequently, we caution against using the dynamical age to make assumptions about
the formation environment of stellar associations. In the best-case scenario,
these groups have high estimated virial ratios; but even then, virial ratios may
be a factor of ten higher than their true values. 

\section{Conclusions}
In this paper, we compare the distribution of stars formed within the \starforge\
simulations with observed stellar associations based on mass, age, and kinematic
measurements in order to derive insights into the origins of these regions and
to constrain the accuracy of derived association properties. Our conclusions are
as follows:

\begin{itemize}
\item   We find that observational estimates of mass and size are highly
        uncertain for true group masses below 100\,\msun\ as these are strongly
        affected by sampling biases caused by observational uncertainties. There
        is wide variation,  between 20 to 90\%, in the recovered mass fraction
        after applying realistic uncertainties. The clusters most affected tend
        to be those that form from initial conditions favouring dynamical
        interactions, such as the densest or lowest virial parameter clouds.
        Groups with higher masses have a mass recovery fraction above 50\%. In
        comparison, velocity dispersion measurements are more reliable and tend
        to be within 40\% of the true value for all group masses.

\item   Most of the missing stellar mass comes from an incomplete census of 
        massive stars (> 6\,\msun), which are commonly saturated in Gaia. In
        \starforge\ stars with masses above 6\,\msun\ are more likely to be
        located in groups, however, this might be an artifact caused by the
        gravitational softening.

\item   For groups with actual virial parameters above 2, dynamical traceback
        can estimate the time since the beginning of the group expansion within
        20\% the true value. However, we find that observational biases can
        cause the virial parameter to be over-estimated by an order of
        magnitude. 

\item   Since stars continuously form during the embedded phase and while 
        expansion occurs, we find little correlation between the later
        association kinematics and the formation timescale. This is true even
        when adopting the true dynamical age rather than the traceback age. 

\item   We match four of the seven distinct associations in Cepheus North 
        with \starforge clusters (CFN-1, CFN-2, CFN-4, and CFN-6). Some regions
        match more than one group in different simulations, while groups in no
        single simulation matched more than two regions simultaneously. The
        observed groups with no match (CFN-3 and CFN-7) still exhibited similar
        velocity dispersions and masses to the other groups, suggesting that
        these two likely formed in similar environments to those  modelled by
        \starforge.

\end{itemize}

In summary, our results emphasize the need for careful consideration of
completeness when addressing the dynamical state of stellar associations. The
kinematical properties of missing members may significantly compromise 
conclusions regarding the formation environment and timescales of stellar
associations.

\section*{Acknowledgements}
We would like to acknowledge the work and trajectory of Sverre Aarseth on
developing the N-Body code that made this work possible. JPF and SSRO are
supported by NASA grant 80NSSC20K0507 and NSF Career award 1748571. We
acknowledge computational resources provided by the University of Texas at
Austin and the Texas Advanced Computing Center (TACC;
http://www.tacc.utexas.edu).

\section*{Data Availability}
The data underlying this article will be shared on reasonable request to the
corresponding author. 


\bibliographystyle{mnras}
\bibliography{references,example} 








\bsp	
\label{lastpage}
\end{document}